\documentclass[preprints,article,accept,moreauthors,pdftex]{Definitions/mdpi}

\firstpage{1} 
\pubvolume{1}
\issuenum{1}
\articlenumber{0}
\pubyear{2021}
\copyrightyear{2020}
\datereceived{} 
\dateaccepted{} 
\datepublished{} 
\hreflink{https://doi.org/} 

\usepackage{lineno,hyperref}
\usepackage{graphicx}
\usepackage{multirow}
\usepackage{mathptmx}
\usepackage{times}
\usepackage{epsfig}
\usepackage{amsmath}
\usepackage{caption}
\usepackage{url}
\usepackage{subfig}
\usepackage{blindtext}
\usepackage{framed,multirow}
\usepackage{booktabs}
\usepackage{soul}
\usepackage{amssymb}
\usepackage{latexsym}
\usepackage{float}

\Title{AWEU-Net: An Attention-Aware Weight Excitation U-Net for Lung Nodule Segmentation}

\TitleCitation{AWEU-Net: An Attention-Aware Weight Excitation U-Net for Lung Nodule Segmentation}

\Author{Syeda Furruka Banu$^{1}$*, 
Md. Mostafa Kamal Sarker $^{2}$, 
Mohamed Abdel-Nasser$^{1,3}$, 
Domenec Puig $^{1}$, 
Hatem  A. Raswan $^{1}$}

\AuthorNames{Syeda Furruka Banu, Md. Mostafa Kamal Sarker, Mohamed Abdel-Nasser, Domenec Puig, Hatem  A. Raswan}

\AuthorCitation{Banu, S.F.; Sarker, M.M.K.;  Abdel-Nasser, M.; Puig, D.; Rashwan, H.A.}

\address{%
$^{1}$ \quad Departament d'Enginyeria Informàtica i Matemàtiques, Universitat Rovira i Virgili, 43007 Tarragona, Spain; syedafurruka.banu@estudiants.urv.cat, mohamed.abdelnasser@urv.cat, domenec.puig@urv.cat, hatem.abdellatif@urv.cat  \\
$^{2}$ \quad National Subsea Centre, Robert Gordon University, Aberdeen, AB10 7GJ, UK; m.kamal.sarker@gmail.com,\\
$^{3}$ \quad Department of Electrical Engineering, Aswan University, Aswan 81542, Egypt; mohamed.abdelnasser@urv.cat}

\corres{Correspondence: syedafurruka.banu@estudiants.urv.cat}

\abstract{Lung cancer is deadly cancer that causes millions of deaths every year around the world. Accurate lung nodule detection and segmentation in computed tomography (CT) images is the most important part of diagnosing lung cancer in the early stage. Most of the existing systems are semi-automated and need to manually select the lung and nodules regions to perform the segmentation task. To address these challenges, we proposed a fully automated end-to-end lung nodule detection and segmentation system based on a deep learning approach. In this paper, we used Optimized Faster R-CNN; a state-of-the-art detection model to detect the lung nodule regions in the CT scans. Furthermore, we proposed an attention-aware weight excitation U-Net, called AWEU-Net, for lung nodule segmentation and boundaries detection. To achieve more accurate nodule segmentation, in AWEU-Net, we proposed position attention-aware weight excitation (PAWE), and channel attention-aware weight excitation (CAWE) blocks to highlight the best aligned spatial and channel features in the input feature maps. The experimental results demonstrate that our proposed model yields a Dice score of $89.79\%$ and $90.35\%$, and an intersection over union (IoU) of $82.34\%$  and $83.21\%$ on the publicly LUNA16 and LIDC-IDRI  datasets, respectively.}


\begin{document}

\section{Introduction}
\begin{figure}[!t]
\centering
\includegraphics[scale=0.6]{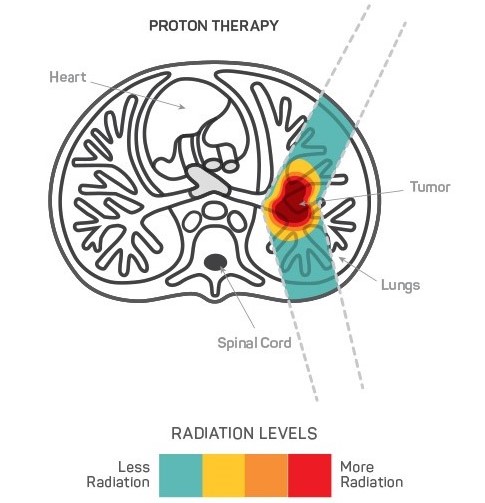}
\caption{A proton therapy plan for lung tumour treatment. The red tumour area is segmented by radiologists (Image from the Seattle Cancer Care Alliance Proton Therapy Center \cite{sccaprotontherapy2020}). }
\label{fig1_photon}
\end{figure}

According to the World Health Organization (WHO), lung cancer is the leading cause of  cancer deaths at 1.80 million in 2020~\cite{cancer2020}.  The estimated new cases rise to 2.89 million with the death projected to reach 2.45 million worldwide by 2030~\cite{factsheets2020}. These deaths could be avoidable by an early diagnosis with a quick treatment plan. The National Lung Screening Trial (NLST) showed that the morality of lung cancer is reduced by 20\% for emphasising the significance of nodule detection and assessment \cite{national2011reduced}. Currently, the efficient investigation to discover pulmonary nodules is based on Computed Tomography (CT) imaging technology that generates hundred images of the lung within a second by a single scan. It is a very difficult and tedious job for radiologists to detect the nodules  from these images manually. However,  computer-aided diagnosis (CAD) systems have assisted radiologists in the automated diagnosis of lung cancer and pulmonary diseases over the last years. These CAD systems mainly depend on the detection and segmentation of various pulmonary parts. They consist of two subsystems, namely;  Computer-aided detection (CADe) and segmentation (CASe), respectively. In the lung cancer screening CAD system, the CADe system identifies the region of interest in the lung nodule and CASe segments the region of the nodule. The detection and segmentation of lung nodules are always challenging tasks because of their heterogeneity in CT images.  However, automated analysis is necessary to measure the properties of the lung nodule for identifying the malignancy of the tumour. The lung nodule segmentation system can determine the malignancy by analysing nodule size, shape and change rate~\cite{callister2015british}. It should be noted that the use of an accurate lung nodule screening CAD system can accelerate the entire diagnosis and radiotherapy process where patients can perform the required radiation or photon therapy (shown in Figure~\ref{fig1_photon}) on the same day.    

To address these premises, in this article we focus on developing a fully automated end-to-end CAD system based on  current state-of-the-art deep learning models.  Many researchers have already handled the lung nodule detection and segmentation problem based on Convolutional Neural Networks (CNNs) and achieved  promising results~\cite{wu2018joint,aresta2019iw,keetha2020u, tang2019nodulenet,cao2020dual,kumar2021lunginfseg,jiang2018multiple}. CNNs  can learn complex features to detect and segment the nodule accurately. However, existing methods are mostly semi-automated and commonly used architectures inspired by U-Net~\cite{ronneberger2015u}. This article proposed AWEU-Net, an attention-aware weight excitation U-Net for lung nodule segmentation. AWEU-Net learns preciously low-level lung nodule-related features using proposed PAWE, and CAWE block. The PAWE and CAWE can enhance the position and channel-based nodule feature representations. The main contributions of this research can be summarized as follows:

\begin{itemize}
    \item [$\bullet$] We present a fully automated end-to-end lung nodule detection and segmentation system.
    \item [$\bullet$] We adopt the Faster R-CNN~\cite{ren2016faster}, a state-of-the-art detection model, improved and named it as ``Optimized Faster R-CNN'' for a reliable lung nodule detection system.
  \item [$\bullet$] We propose AWEU-Net, an efficient  fully automated lung nodule segmentation model.
 \item [$\bullet$] We propose the use of PAWE and CWEU mechanisms to discover the correlation between the position and channel features and enhance the model ability to distinguish between lung nodule and normal tissue feature representations.
 \item [$\bullet$] We assess The AWEU-Net model on two publicly datasets; LUNA16 and LIDC-IDRI and demonstrate that its performance overcomes the current state-of-the-art models.
 \end{itemize}

This article is prepared as follows. Section~\ref{related_work} reviews the existing lung nodule segmentation systems based on classical computer vision and deep learning techniques. The proposed system workflow and model architecture are explained in Section~\ref{proposed_methodology}. The experimental results are demonstrated in Section~\ref{exp_results}. Finally, Section~\ref{conclusions} concludes the article with some future lines of this research. 

\section{Related work}
\label{related_work}
During the last decade, several lung nodule detection and segmentation systems based on classical computer vision and deep learning techniques were presented.  The most common lung nodule segmentation techniques are discussed in this section and summarized in Table~\ref{table1}.
\begin{table}[!h]
\centering
\caption{Summary of existing lung nodule segmentation methods. The undeclared information is marked with dashes (-) in the referred literature.}
\label{table1}
\scalebox{0.60}{%
    \begin{tabular}{@{}l|l|l|l|l@{}}
    \toprule \toprule
    References           & Methods/Architectures                                                                                             & Dataset                & Pre-Processing                                                & Post-Processing          \\ \midrule \midrule
    \multicolumn{5}{l}{\textit{Classical computer vision-based}}                                                                                                                                                                                                          \\ \midrule 
    \qquad \cite{dehmeshki2008segmentation}        & Region growing                            & PRIVATE              & Local Contrast \& Hole Filling                    & - \\\hline
    \qquad \cite{tan2013segmentation}        & Active contours                            & LIDC-IDRI              & Thresholding \& morphological operations                     & Markov random field \\\hline
    \qquad \cite{farag2013novel}        & Level sets                            & LIDC-IDRI              & Statistical intensity                    & Region condition \\\hline
    \qquad \cite{dai2015novel}        & Graph cuts                           & PRIVATE              & Gaussian smoothing                    & - \\\hline
    \qquad \cite{navya2018lung}        & Adaptive thresholding                            & LIDC-IDRI              & Histogram equalization \& noise filtering                   & Morphological operations \\\hline
    \qquad \cite{li2020segmentation}        &  GMM fuzzy C-means                           & LIDC-IDRI \& GHGZMCPLA              & Non-local mean filter \& gaussian pyramid                   & Random walker \\\hline
    \qquad \cite{savic2021lung}        &  Region-based fast marching                          & LIDC-IDRI             & Convex hulls                   & Mean threshold \\ \midrule

    \multicolumn{5}{l}{\textit{Deep learning-based}}                                       
    
    \\\midrule 

\qquad \cite{wu2018joint}        &  U-Net                          & LIDC-IDRI               & Nodule ROI selection                  & -\\\hline

\qquad \cite{aresta2019iw}        &  iW-Net                          & LIDC-IDRI             & Nodule ROI selection                  & -\\\hline

\qquad \cite{keetha2020u}        &  U-Det                         &  LUNA16              & Data augmentation                   & -\\\hline

\qquad \cite{tang2019nodulenet}        &  Nodulenet                         &  LIDC-IDRI \& LUNA16              & Nodule ROI selection \& Data augmentation                   & -\\\hline

\qquad \cite{cao2020dual}        &  DB-ResNet                         &  LIDC-IDRI              & Nodule ROI selection \& Data augmentation                   & Remove the noisy voxel \\\hline

\qquad \cite{jiang2018multiple}        &  MRRN                        &  TCIA \& MSKCC \& LIDC-IDRI              & Nodule ROI selection \& Data augmentation                   & - \\ \bottomrule \bottomrule
\end{tabular}%
}
\end{table}

\subsection{Traditional computer vision-based approaches}
In the field of lung nodule analysis, many computer vision methods have been used like region growing~\cite{dehmeshki2008segmentation}, active contours~\cite{tan2013segmentation}, level sets~\cite{farag2013novel}, graph cuts~\cite{dai2015novel}, adaptive thresholding~\cite{navya2018lung},  Gaussian mixture model (GMM) with fuzzy C-means~\cite{li2020segmentation}, and region-based fast marching~\cite{savic2021lung}. For instance, a contrast-based region growing method and fuzzy connectivity map of the object of interest were used  in~\cite{dehmeshki2008segmentation}  to segment the various types of pulmonary nodules. This method did not perform well with irregular nodules because of the merging criterion in the region growing technique that needs accurate settings. A geometric active contours with a marker-controlled watershed as well as Markov random field (MRF) was used in~\cite{tan2013segmentation} to segment the lung nodule. This method depends on manually selects a region of interest (ROI) in the nodule region.  ~\cite{farag2013novel} used the shape prior hypothesis along with level sets that iteratively minimizes the energy to segment the juxtapleural nodules. The precision of this method also depends on the selection of the ROIs. A graph cuts with the expectation-maximization (EM) algorithm was proposed in ~\cite{dai2015novel} for lung segmentation on chest CT images. This algorithm has a high computational cost because its main focus is on the Gaussian mixture model(GMM) training method and the creation of the corresponding graph. ~\cite{navya2018lung} used the adaptive thresholding along with watershed transform to detect the nodules. This approach mainly relies on several image pre- and post-processing procedures. ~\cite{li2020segmentation} combined GMM earlier knowledge within the conventional fuzzy C-means method to improve the robustness of pulmonary nodules segmentation. The major disadvantage of fuzzy C-means algorithms is that they are sensitive to noise, outliers and primary cluster selection. A region-based approach was introduced in~\cite{savic2021lung} by using the fast marching method, which gives a precise segmentation of the nodule and can properly handle juxtapleural and juxtavascular nodules. All these above-mentioned traditional  approaches are semi-automated or depend on several image pre- and post-processing methods. 

\subsection{Deep learning-based approaches}
Recently, many researchers have developed various deep learning-based systems for lung nodule detection and segmentation. \cite{wu2018joint} presented a simple U-Net for for lung nodule  segmentation  by utilizing half of the convolutional layers normally used in the original U-Net~\cite{ronneberger2015u}. The model has many window widths and window centres enhancing the nodule features. It improves the model performance over the original U-Net by 2\% in Dice Score Coefficient (DSC). ~\cite{aresta2019iw} combined two U-Net models (named iW-Net)  that can be used with and without user interaction. In the first instance, the user selects the nodules ROI and the corresponding end-point to produce a weight map for developing the prediction of the model. The architecture is designed by considering the expected round shape of the nodules. The loss function is also combined with the weight map and the feature of the model output. \cite{jiang2018multiple} presented a multiple resolution residual network (MRRN) that is a modification of the ResNet~\cite{he2016deep} based on the U-Net model.  A slightly transformed version of U-Net called U-Det was presented in~\cite{keetha2020u}, where many hidden layers were used to filter the residuals blocks located within the encoder and decoder. U-Det also applies the Mish activation function. An end-to-end 3D Deep-CNN called NoduleNet  was presented in~\cite{tang2019nodulenet} for detection and segmentation jointly of the pulmonary nodule. NoduleNet uses an UNet-like model to detect the nodule and then runs a segmentation distillation  on the Volume of Interest (VOI) surrounding the detected nodule, gradually up-sampled the segment volumes and integrates them with low-level features. NoduleNet solves the loss of resolution inside the VoI by duplicating the pooling layers and the convolutions of the image. ~\cite{cao2020dual} presented a dual-branch residual network (DB-ResNet) that achieved results similar to~\cite{wu2018joint}. The major differences between ~\cite{cao2020dual} and ~\cite{wu2018joint} are the use of residuals layers, two slightly modified pooling layers and the convolutional blocks of ResNet~\cite{he2016deep}. 

\section{Proposed methodology}
\label{proposed_methodology}
\begin{figure}[!t]
\centering
\includegraphics[width=0.7\textwidth]{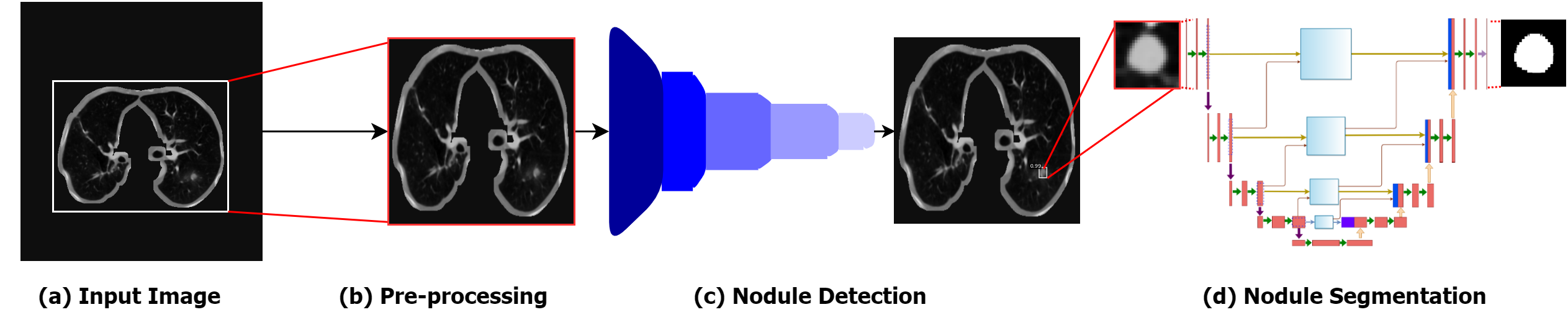}
\caption{The step-by-step workflow of the proposed method. (a) The converted input image is extracted from the original CT slice (b) The pre-processing step for selecting ROI of lung, (c) The detection of lung nodule using Optimized Faster R-CNN, and (d) The segmentation of lung nodule using the proposed AWEU-Net model.}
\label{fig2_workflow}
\end{figure}

The main workflow of the   proposed method is illustrated in Figure~\ref{fig2_workflow}. As a pre-processing stage, we extracted the CT volumes slice by slice. The extraction of the CT slices are converted into images from the original CT scans (``.dcm'' file). To detect the nodule ROI, the Optimized Faster R-CNN method is then used. The detected nodule ROI is fed as an input to the segmentation task. The proposed AWEU-Net is used to segment the nodule preciously, as detailed below.

\subsection{Pre-processing}
The raw CT scans data are available in the ``.dcm'' format which is processed with the pylidc~\cite{hancock2016lung} library to convert them to the ``.png''  format for the purpose of rendering it more meaningful and useful as shown in Figure~\ref{fig2_workflow}-(a). Afterwards, a set of image processing techniques (Otsu with a binary threshold and morphological dilation) are applied to obtain  the lung region from the converted image as shown in Figure~\ref{fig2_workflow}-(b). We choose the size of the extracted lung region to be $512\times512$ pixels. We split the data set into 70\% for training,  10\% for validation, and 20\% for test. We convert all ground-truth images to ms-coco format~\cite{lin2014microsoft} for using the Optimized Faster R-CNN to detect the lung nodules. 

\begin{figure}[!b]
\centering
\includegraphics[width=0.7\textwidth]{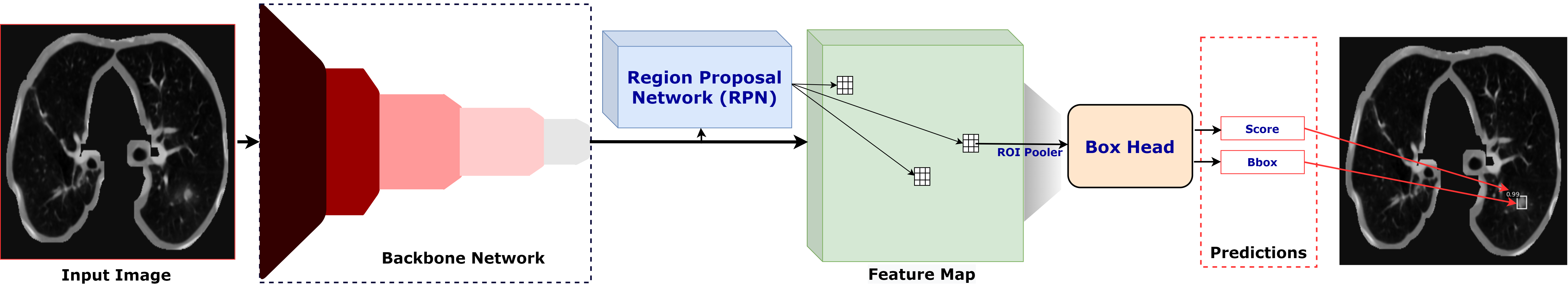}
\caption{The details architecture of Optimized  Faster R-CNN.}
\label{figure_nodule_detection}
\end{figure}

\subsection{Nodule detection model}
In the nodule detection model, we used Optimized Faster R-CNN based on the original Faster R-CNN~\cite{ren2016faster} to automatically detect lung nodules based on the lung images as shown in Figure~\ref{fig2_workflow}-(c). The model is a  two-stage network with three main blocks, Backbone Network, Region Proposal Network (RPN), and Box head, as shown in Figure~\ref{figure_nodule_detection}. We used ResNet50~\cite{he2016deep} as a backbone network to extract feature maps from the input image. The feature map is then fed into the RPN to perform boundary regression and classification analysis. The classification principle is based on which a candidate frame is either related to background or to the object. The position and score of the RPN outputs on the candidate frame are sent to the Box head, where the final regression and classification of the object is performed. Finally, the prediction will show the bounding box of the target (nodule) with the classification score. 

\subsection{Nodule segmentation model}
\begin{figure}[!t]
\centering
\includegraphics[width=0.6\textwidth]{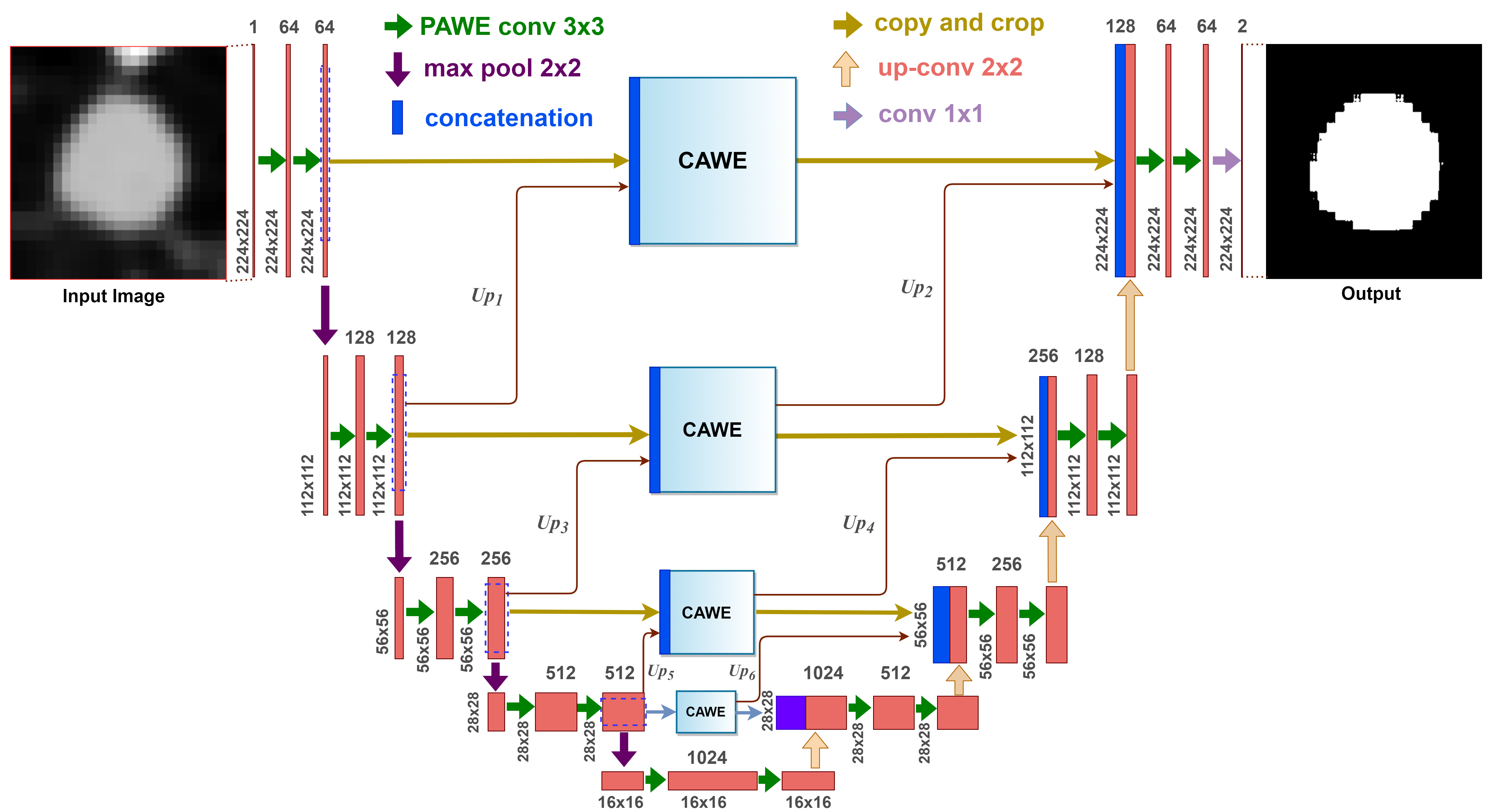}
\caption{The  architecture of the proposed AWEU-Net. The PAWE and CAWE block refers the position attention-aware weight excitation and channel attention-aware weight excitation, respectively.}
\label{fig3_propsed_net}
\end{figure}

 We crop the ROI based on the detection box suggested by the nodule detection model introduced above. We  resize the ROI to $224\times224$ and feed it into the proposed nodule segmentation model. We propose an attention-aware weight excitation U-Net, AWEU-Net, for our lung nodule segmentation  (Figure~\ref{fig3_propsed_net}). The network is based on the U-Net~\cite{ronneberger2015u}, which is a well-known deep learning model for medical image segmentation. The AWEU-Net model learns to segment the input sub-images by determining the boundaries of the nodule region to discriminate between normal and abnormal tissues. One of the main contributions in this article is to develop a PAWE block in the AWEU-Net model to capture the contextual positional features of the input image. We also propose another CAWE block to enhance the channel-wise feature maps that are coming from each layer of the AWEU-Net model. The details about PAWE and CAWE will be discussed in Section~\ref{pawe} and Section~\ref{cawe}, respectively.

The AWEU-Net architecture is composed of two successive networks: encoder and decoder.  The encoder consists of a sequence of  PAWE  blocks and max-pooling layers. Each encoder layer is composed of a PAWE  block with a $3\times3$ convolutional layer followed by a ReLU as an activation function. Four down-sample blocks with $2\times2$ max-pooling followed by a stride of $2$ are used after each block of the encoder. The decoder consists of a sequence of up-convolutions and concatenation with the corresponding high-resolution features from the CAWE blocks that provide a high-resolution output segmentation map. The features coming from encoder layers are also upsampled (\textit{Up1}, \textit{Up3} and \textit{Up5}) and concatenated with CAWE blocks. The same features coming from CAWE blocks also upsampled (\textit{Up2}, \textit{Up4}, and \textit{Up6}) and concatenated with corresponding decoder layers.  The decoder network consists of four layers similar to the encoder. Each layer also consists of a PAWE  block with a $3\times3$ convolutional layer followed by a ReLU and a $2\times2$ up-conv layer.  After each decoder layer, the feature maps are upsampled to the same size of the CAWE block output to keep the consistency and concatenate it. This mechanism enhances the positional and channel attention-based features learned from the encoder phase and utilises them for the reconstruction means in the decoder network. The final output layer of the model applies a $1\times1$ convolutional layer to map the final 64 feature vector to the number of targeted segmentation classes. It should be noted that in our case the segmentation classes are the background and lung nodule (two classes).
      
\subsubsection{Position attention-aware weight excitation (PAWE)}
\label{pawe}
\begin{figure}[!t]
\centering
\includegraphics[width=0.6\textwidth]{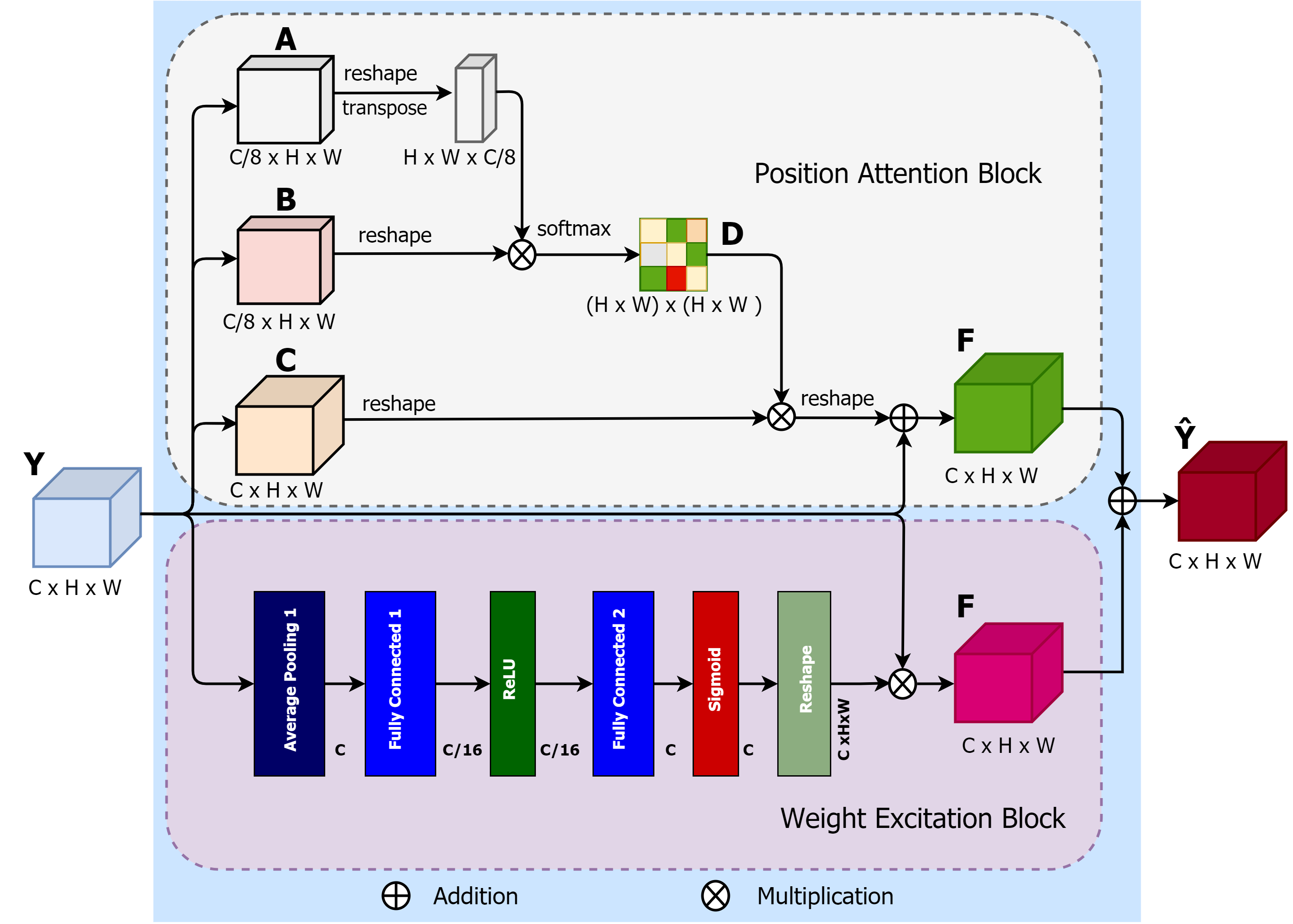}
\caption{Illustration of proposed PAWE block.}
\label{fig4_propsed_pawe_block}
\end{figure}

The PAWE block consists of two sub-blocks: the position attention block (PAB) and the weight excitation block (WEB). To demonstrate the proposed PAWE block, let the input feature is $Y \in \mathbb{R}^{C\times H\times W}$, where $C$, $H$ and $W$ are channel, height and width, respectively (see Figure~\ref{fig4_propsed_pawe_block}). In the PAB block, $Y$ is fed into three convolution layers that produces three new feature maps $A$, $B$ and $C$, respectively. The produced feature maps, $A^p, B^p\in \mathbb{R}^{C/8\times H\times W}$  are from the first two convolutional layers, where $p$ superscript is for PAB.Then, $A^p$ and $B^p$ feature maps are reshaped into $(H\times W)\times C/8$. A matrix multiplication is implemented between the transpose of $\mathbf{A^p}$ and $\mathbf{B^p}$, and a spatial attention map is produced, $D^p \in \mathbb{R}^{(H \times W) \times (H \times W)}$ by using a softmax function:
\begin{equation}
    d_{i,j}^p = \frac{\exp{(A{i}^p \cdot B{j}^p)}}{\sum_{i=1}^{H \times W}\exp{(A{i}^p \cdot B{j}^p)}},
    \label{eq1}
\end{equation} 
where $s_{i,j}$ indicates the ${i^{th}}$ position's associated on position of ${j^{th}}$. The softmax function $D^p$ attempts to learn the relationship between two spatial positions in the input feature maps. 

In addition, the output of the third convolutional layer  
$C^p \in \mathbb{R}^{C\times (H\times W)}$ is also reshaped to the same shape of the input feature map $Y$ and then multiplied by a permuted order of the spatial attention map $D^p$ of (\ref{eq1}). The final output is reshaped to a $\mathbb{R}^{C\times (H\times W)}$ to provide the final feature map of PAB block, $F$ as,  
\begin{equation}
    F_{PAB,j} = \alpha_p\sum_{i=1}^{H\times W}s_{ij}^pC{j}^p + Y_j,
    \label{equF1}
\end{equation}
where $\alpha_p$ is defined as $0$ as explained in~\cite{fu2019dual}. The resulting feature $\mathbf{F}$ at every position is a weighted sum of the entire neighbours of original features. Note that all ``$^p$'' notations are defined for the position.  
 
In the WEB,a sub-network for the location-based weight excitation (LWE) proposed in~\cite{quader2020weight} is used. The LWE provides fine-grained weight-wise attention during the backpropagation. The WEB shown in (Figure~\ref{fig4_propsed_pawe_block}) is defined as:

\begin{equation}
    F_{WEB,j} = Re2(FC2(Re1(FC1(AP(W_{WEB,j}))))),
    \label{equ3}
\end{equation}
where $W_{WEB,j}$ is the weights across the $j^{th}$ output channel. The average pooling layer, $AP$, averages the values of each  $H \times W$. $Re1$ and $Re2$ are two ReLu activation functions.  $FC1$ and $FC2$ are two fully connected layers. 
 
The output feature from WEB is reshaped and multiplied to the input feature map. Finally, an element-wise sum operation is performed between the features maps from the PAB and WEB to produce the final PAWE features as follows:
\begin{equation}
    \hat{Y}_{PAWE,j} = F_{PAB,j} +  F_{WEB,j},
    \label{equ4}
\end{equation}
This process generates a global contextual description and aggregates the context according to a spatial weighted attention map by creating relevant weighted features that can produce common weight-excitation and enhance the intra-class semantic coherence.

\subsubsection{Channel attention-aware weight excitation (CAWE)}
\label{cawe}
Like PAWE, the proposed CAWE block includes two sub-blocks, a channel attention block (CAB) and a weight excitation block (WEB). In the CAB block, the input $Y \in \mathbb{R}^{C\times H\times W}$ is reshaped in the initial two steps and permuted in the second part into  $Y_{1}^c \in \mathbb{R}^{(H\times W)\times C}$ and $Y_{2}^c \in \mathbb{R}^{C \times (H\times W)}$, where the superscript $c$ is defined for CAB. Afterwards, a matrix multiplication between $Y_{1}^c$ and $Y_{2}^c$ is performed. The channel attention map $E^c \in \mathbb{R}^{C \times C}$ is defined as: 
\begin{equation}
    e_{i,j}^c = \frac{\exp{(Y_{1,i}^c \cdot Y_{2,j}^c)}}{\sum_{i=1}^{C}\exp{(Y_{1,i}^c \cdot Y_{2,j}^c)}},
\end{equation}
where the outcome of the $i^{th}$ channel on the $j^{th}$ is produced by $e_{i,j}^c$. 
A multiplication of transposed version of the input feature maps, $Y_{3}^c$ reshaped to $\mathbb{R}^{C \times (H\times W)}$, and the resulted $E^c$ is performed. Consequently, the final channel attention map can be defined as:
\begin{equation}
    F_{CAB,j} = \alpha_c\sum_{i=1}^{C}e_{ij}^cY_{3,j}^c + Y_j,
\end{equation}
where $\alpha_c$ quantify the weight of the channel attention map by the input feature map $Y$. The final WEB sub-network feature map can be obtained from the Equation~\ref{equ3}. 

Finally an element-wise sum operation is performed between the CAB and WEB output features maps to produce the final CAWE features as follows:
\begin{equation}
    \hat{Y}_{CAWE,j} = F_{CAB,j} +  F_{WEB,j}  ,
\end{equation}
This process emphasizes class-dependent feature maps  using weighted excitation versions of the features of all the channels and boosting the feature difference among the classes. Note that all ``$^c$'' notations are defined for the channel. 

\begin{figure}[!t]
\centering
\includegraphics[width=0.6\textwidth]{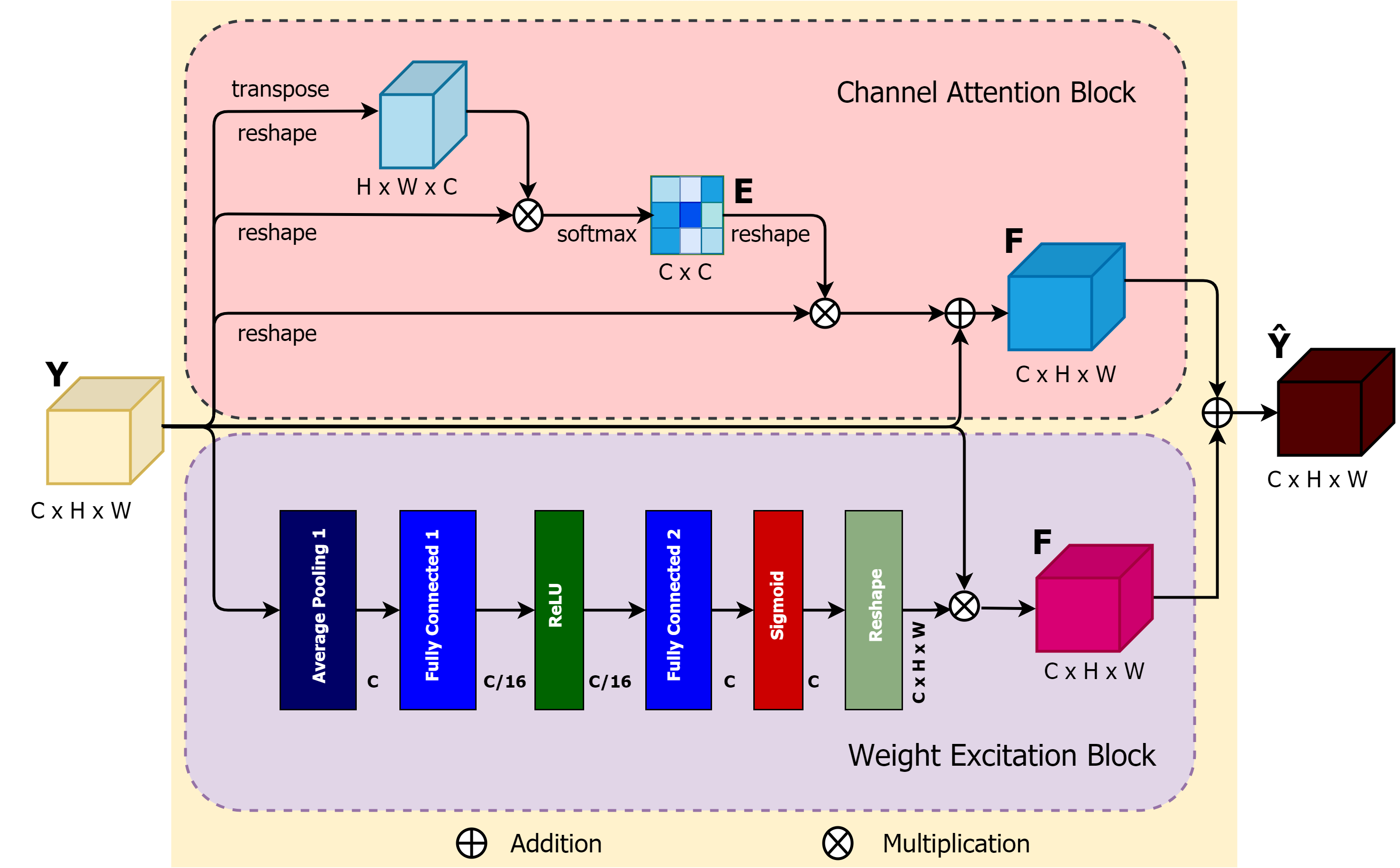}
\caption{Illustration of the proposed CAWE block.}
\label{fig5_propsed_cawe_block}
\end{figure}

\section{Experimental results and discussion}
\label{exp_results}

\subsection{Datasets}
\label{dataset}
In this work, we used two publicly available datasets:
\begin{itemize}
    \item Lung Image Database Consortium image collection (LIDC-IDRI)~\cite{armato2011lung}  consists of 1018 CT scans performed on 1010 patients from seven different organisations. Each CT scan has been analysed by four radiologists, who individually identified the nodule and manually segmented the region of all the nodules with a diameter larger than three millimetres. Each CT scan can include one or more nodule regions, so the total segmented masks are 5066. Looking close at the dataset, many nodules are very small and not satisfied the malignancy index. Therefore, we used a diameter threshold larger than 20 mm to excluded all tiny nodules from our dataset. Afterwards, we split our final dataset, which contains 2044 nodule masks in total, into  train, validation and test sets of $70\%$, $10\%$, and $20\%$ respectively.
    \item LUng Nodule Analysis 2016 (LUNA16)~\cite{setio2017validation}  is derived from the LIDC-IDRI dataset~\cite{armato2011lung}. It contains  888 CT scans from the LIDC-IDRI for the grand challenge with round annotation masks for all the segmented nodules. LUNA16 challenge dataset contains 1186 nodules annotations. We obtained 2300 nodule masks from the annotation after pre-processing. We split the dataset into train, validation and test sets similar to the LIDC-IDRI dataset.
\end{itemize}

\subsection{Model Implementation}
We individually trained the nodule detection and segmentation models on the PyTorch framework~\cite{paszke2019pytorch}. To train the detection model, the Stochastic Gradient Descent (SGD)~\cite{gulcehre2017robust} optimizer with a learning rate of $0.002$ was used. The Binary Cross-Entropy (BCE) and the $L1$ norm loss functions were used to train the detection model with a batch size of $4$. On the other hand, the Adam~\cite{kingma2014adam} optimizer with a learning rate of $0.0002$, the BCE and the IoU loss functions were also used to train the segmentation model with a batch size of $4$. Note that, data-augmentation was applied during training for both detection and segmentation models to increase the size of the training dataset. We augmented the datasets by random rotation, flipping horizontally and vertically and applying the elastic transform. Finally, all the experiments were carried on NVIDIA GeForce GTX 1080 GPU with $8GB$ memory and running about $10-15$ hours to train $100$ epochs for each model.

\subsection{Model Evaluation}
Two different procedures were used on both datasets to evaluate the proposed detection and segmentation models. For Pixel-level evaluation, the segmentation model provides a pixel-wise output of the class probabilities for every pixel in the input nodule ROIs. The output is converted into a binary segmentation map using a threshold value. Regarding pixel-level evaluation metrics, accuracy (ACC), sensitivity (SEN) and specificity (SPE) are calculated to evaluate the performance of the segmentation model. We also plot a receiver operating characteristic (ROC) curve for calculating Area Under the curve (AUC). For Object-level evaluation, we used the segmentation output to calculate the Dice coefficient (DSC) and intersection over union (IoU) for assessing the ability of the algorithm to preciously segment the boundaries of the nodule. Note that in our case, there is no ``true negative'' class, since there is no ``object'' corresponding to the absence of nodules. Besides, we also plot the precision-recall (PR) curve instead of the ROC to compare the ground truth number and find the correlation.

\begin{table}[!h]
\centering
\caption{The average precision (AP) comparison of the four detection models.}
\label{table_detection_results}
\resizebox{0.35\textwidth}{!}{%
\begin{tabular}{@{}l|l|l@{}}
\toprule
Datasets & Models & AP(\%) \\ \midrule
\multirow{4}{*}{LIDC-IDRI} & Optimized Faster R-CNN & \textbf{91.44} \\
 & Original Faster R-CNN & 85.45 \\
 & Fast R-CNN & 79.41 \\
 & R-CNN & 75.48 \\ \midrule
\multirow{4}{*}{LUNA16} & Optimized Faster R-CNN & \textbf{92.67} \\ 
 & Original Faster R-CNN & 89.31 \\
 & Fast R-CNN & 82.32 \\
 & R-CNN & 78.17 \\ \midrule
\end{tabular}%
}
\end{table}

\subsection{Nodule Detection}
\begin{figure}[!t]
\centering
\includegraphics[width=0.6\textwidth]{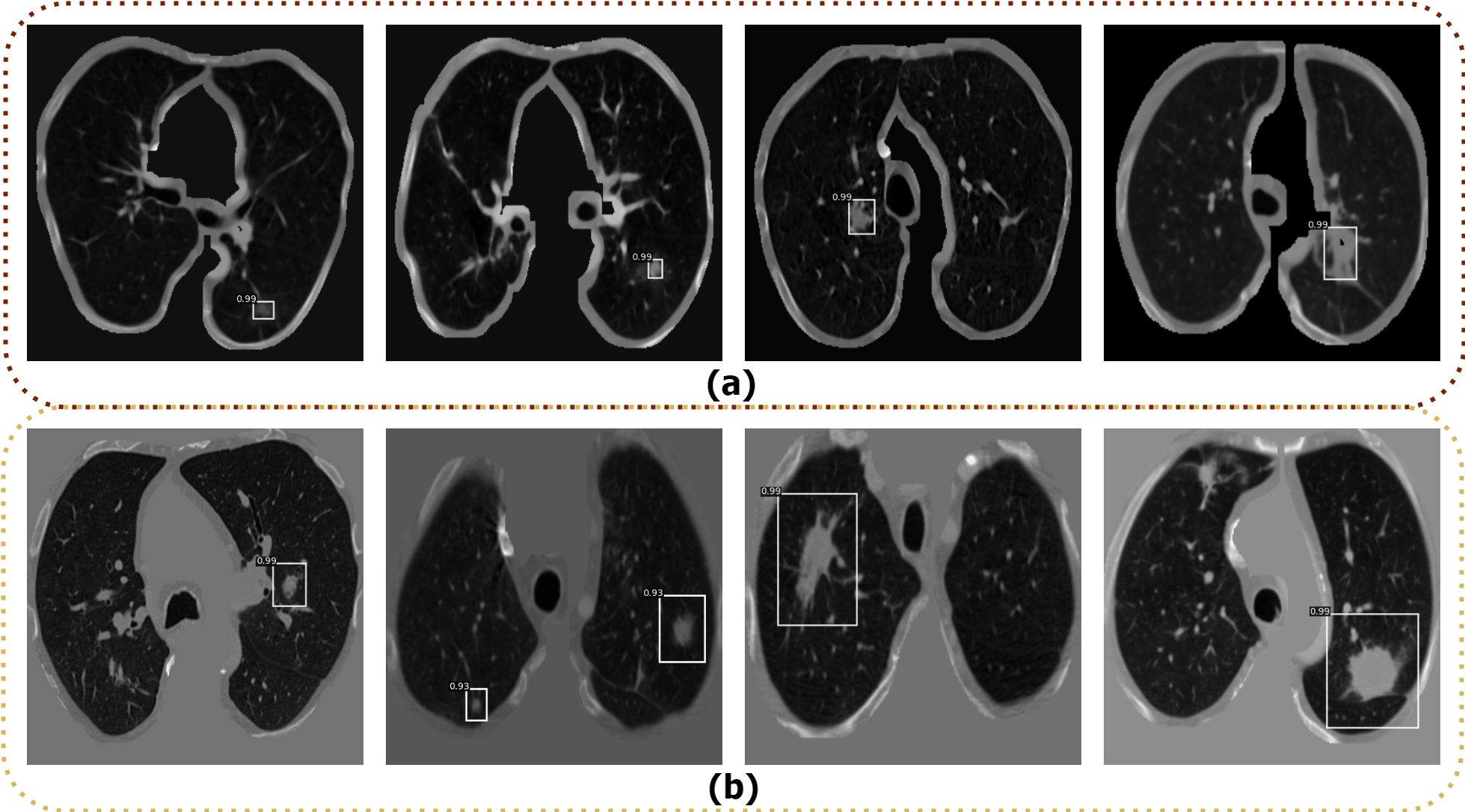}
\caption{Examples of some lung nodule detection using Optimized Faster R-CNN; (a) Detection results from LIDC-IDRI dataset (b) Detection results from LUNA16 dataset.}
\label{fig_detection}
\end{figure}

To detect the nodule in the input CT images, we used different state-of-the-art deep learning detectors models, such as R-CNN~\cite{girshick2014rich}, Fast-RCNN~\cite{girshick2015fast}, original Faster R-CNN~\cite{ren2016faster} and Optimized Faster R-CNN. The aforementioned detection models were trained and tested on LIDC-IDRI and LUNA16 datasets. To training the above models, we used the data splits as discussed in Section~\ref{dataset}. We used all default parameters for training the R-CNN~\cite{girshick2014rich}, Fast-RCNN~\cite{girshick2015fast}, and original Faster R-CNN~\cite{ren2016faster} models based on their original paper. We fine-tune the parameters of the original Faster R-CNN to find the best parameters to achieved the highest performance and named it Optimized Faster R-CNN. The best combination for this model is with a learning rate of 0.001, step size of 70000, gamma of 0.1, and the dropout ratio of 0.5. The model was trained by the pre-trained ResNet50 model to extract the features with a batch size of 64. We finally compare the average precision (AP) of the detection as shown in Table~\ref{table_detection_results} to select the best detection model among the tested models. The Optimized Faster R-CNN model yields the best results, with the highest AP on both datasets. In turn, R-CNN, Fast R-CNN, original Faster R-CNN models have not properly detected all nodules in the input CT images. Therefore, we have selected the Optimized Faster R-CNN model to detect the nodule in CT images. Some examples of lung nodule detection using Optimized Faster R-CNN are shown in Figure~\ref{fig_detection}. As shown, the Optimized Faster R-CNN model is able to detect the nodule regions even the small nodules.

\subsection{Nodule Segmentation}
The proposed lung nodule segmentation model is compared to the state-of-the-art approaches and evaluated in terms of quantitative and qualitative results. For the quantitative study, we used ACC, SEN, SPE for pixels-level and DSC and IoU for object-level performance, respectively as shown in Table~\ref{table_segmentaion_results}. We compared the AWEU-Net to six different lung nodule segmentation models considering both datasets:  PSPNet~\cite{zhao2017pyramid}, MANet~\cite{fan2020ma}, PAN~\cite{li2018pyramid}, FPN~\cite{lin2017feature}, DeeplabV3~\cite{chen2017rethinking}, and U-Net~\cite{ronneberger2015u}. As shown in Table~\ref{table_segmentaion_results}, AWEU-Net outperforms all the tested models in terms of the ACC, SPE, DSC, and IoU metrics on the LUNA16 dataset. AWEU-Net yields ACC, SPE, DSC, and IoU scores of 91.32\%, 93.46\%, 89.79\%, 82.32\%, and 89.88\%, respectively  which is  1.18\%, 1.47\%, 0.97\%, 1.8\%, and 0.93\%  points higher than the scores of the second-best method (i.e., U-Net). In turn, the DeeplabV3 achieved a SEN score of 93.01\%, which is 1.32\% point higher than AWEU-Net. However, the proposed segmentation model provides a comparable SEN score of $91.69\%$.

\begin{table}[!h]
\centering
\caption{Comparison between the proposed AWEU-Net and other six models on the LIDC-IDRI and LUNA16 test datasets (bold represent the best performance).}
\label{table_segmentaion_results}
\resizebox{0.6\textwidth}{!}{%
\begin{tabular}{@{}l|l|l|l|l|l|l@{}}
\toprule
Datasets                   & Models             & ACC                        & SEN                        & SPE                        & DSC                        & IoU                                         \\ \midrule
\multirow{7}{*}{LUNA16}    & PSPNet             & 0.8718                     & 0.8711                     & 0.9012                     & 0.8512                     & 0.7513                                        \\
                           & MANet              & 0.8874                     & 0.8686                     & 0.9285                     & 0.8663                     & 0.7743                                        \\
                           & PAN                & 0.8604                     & 0.8709                     & 0.8873                     & 0.8424                     & 0.7354                                       \\
                           & FPN                & 0.8846                     & 0.9143                     & 0.8905                     & 0.8722                     & 0.7806                                      \\
                           & DeeplabV3          & 0.8918                     & \textbf{0.9301}                     & 0.8910                     & 0.8794                     & 0.7916                                        \\
                           & U-Net              & 0.9014                     & 0.9136                     & 0.9199                     & 0.8882                     & 0.8054                                         \\
                           & \textbf{Proposed\_AWEU-Net} & \textbf{0.9132}                     & 0.9169                     & \textbf{0.9346}                     & \textbf{0.8979}                     & \textbf{0.8234}                                      \\\midrule 
\multirow{7}{*}{LIDC-IDRI} & PSPNet             & 0.9309                     & 0.8514                     & 0.9620                     & 0.8684                     & 0.7783                                       \\
                           & MANet              & {0.9327} & {0.8749} & {0.9557} & {0.8788} & {0.7905}  \\
                           & PAN                & {0.9268} & {0.8369} & {0.9603} & {0.8577} & {0.7653} \\
                           & FPN                & {0.9393} & {0.8981} & {0.9562} & {0.8934} & {0.8127} \\
                           & DeeplabV3          & {0.9429} & {0.9023} & {0.9602} & {0.8983} & {0.8191}  \\
                           & U-Net              & {0.9436} & {0.8968} & {0.9635} & {0.8987} & {0.8200}  \\
                           & \textbf{Proposed\_AWEU-Net} &   \textbf{0.9466}                             &    \textbf{0.9084}                              &  \textbf{0.9641}                              &        \textbf{0.9035}                       &           \textbf{0.8321}                                       \\ \midrule 
\end{tabular}%
}
\end{table}

\begin{figure}[!ht]
\centering
\includegraphics[trim={0 0 0 0},clip,scale=.15]{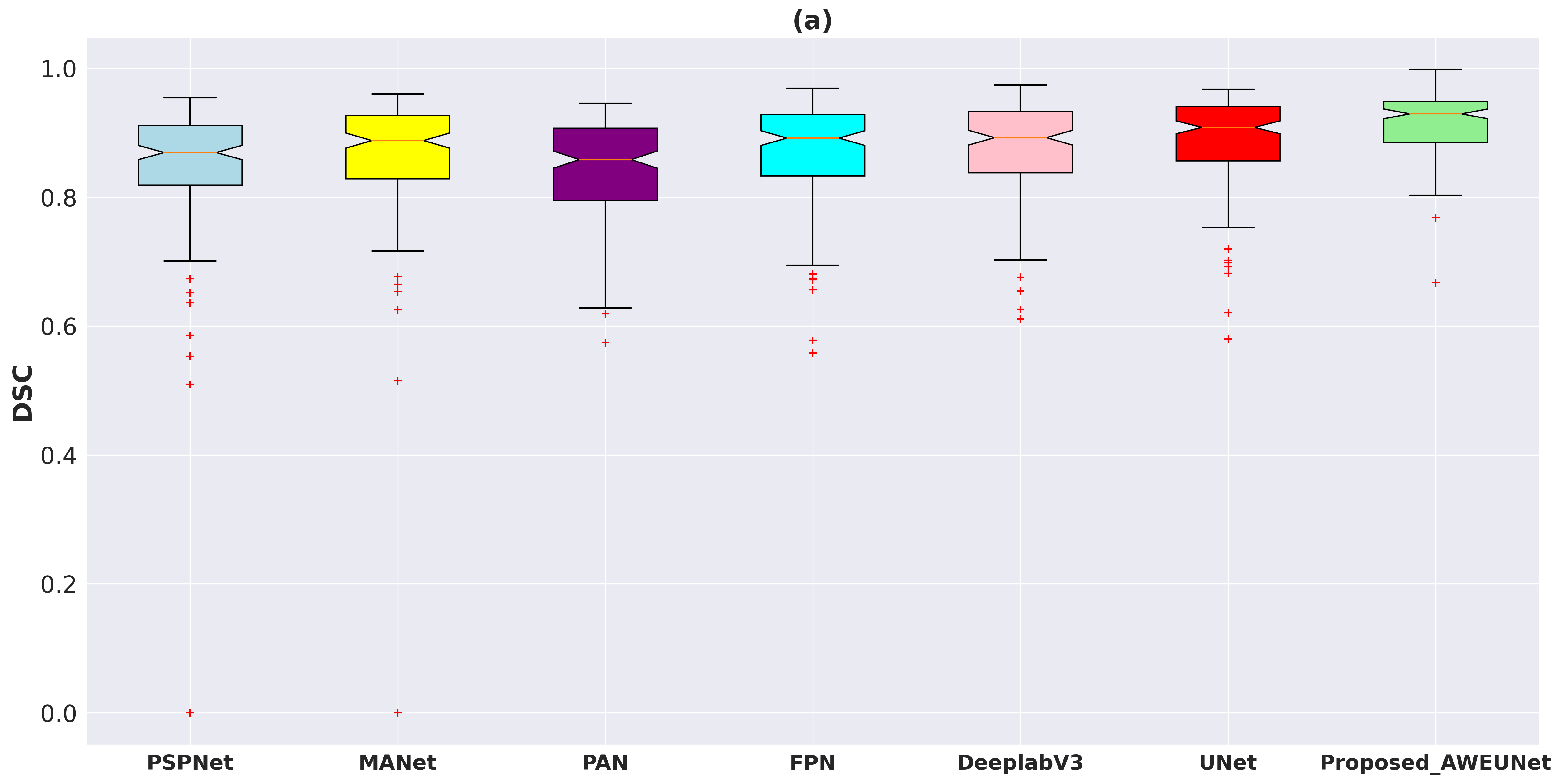}
\includegraphics[trim={0 0 0 0},clip, scale=.15]{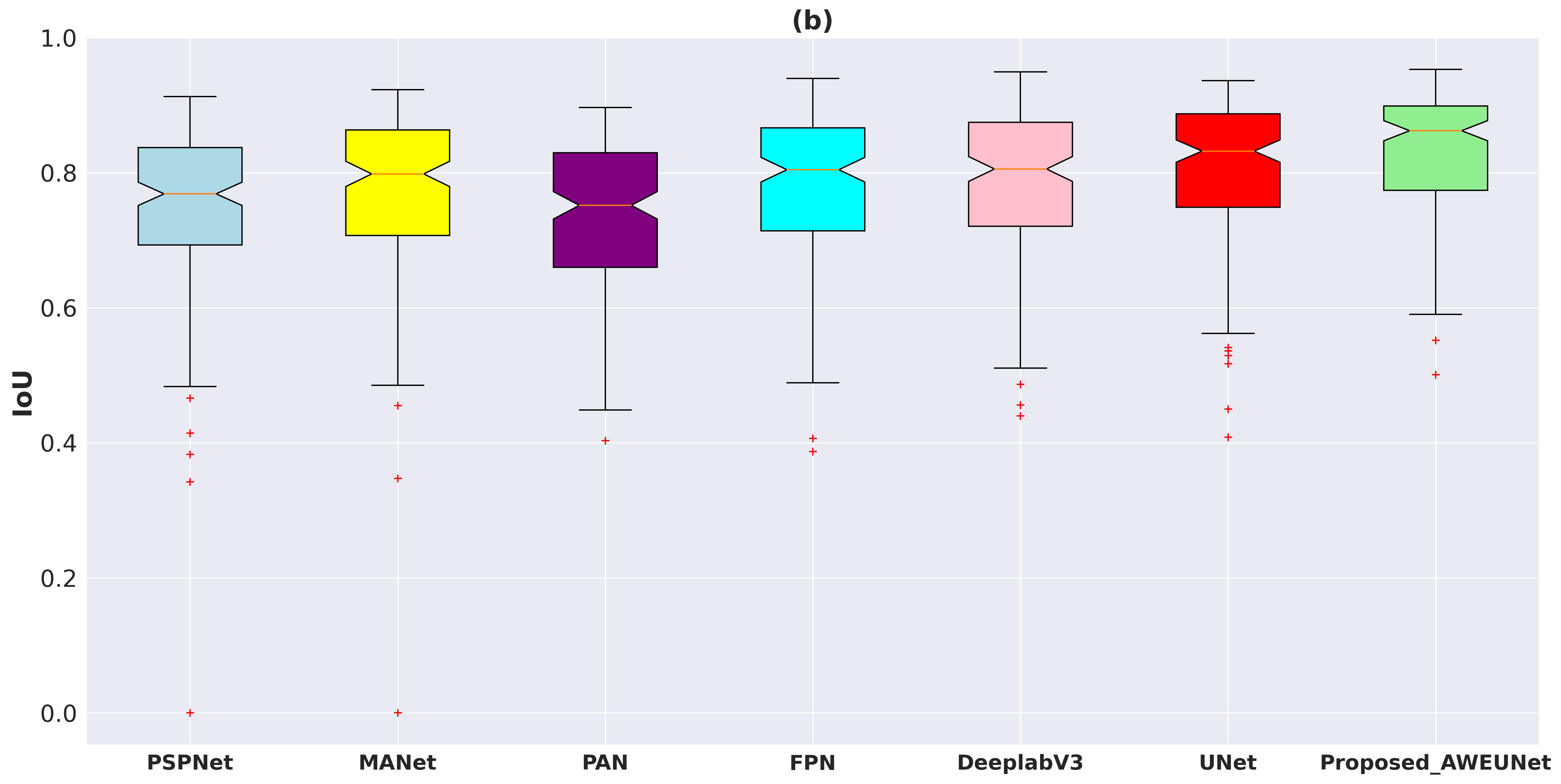}
\caption{Boxplots of (a) Dice coefficient (DSC) and (b) Intersection over Union (IoU) scores for all test samples of LUNA16 lung nodule segmentation dataset. Different boxes indicate the score ranges of several methods; the red line inside each box represents the median value; and all values outside the whiskers are considered as outliers, which are marked with the (+) symbol.}
\label{fig_dsc_iou_luna16}
\end{figure}

In addition, using the test set of the LUNA16 and LIDC-IDRI datasets, the box plots of DSC and IoU scores of the six models and AWEU-Net were drawn to demonstrate the segmentation ability of AWEU-Net as shown in Figure~\ref{fig_dsc_iou_luna16}. On both datsets, the proposed AWEU-Net yields the higher DSC and IoU mean scores and the lowest standard deviation with only two outliers compared to the other six segmentation models that are represented many outliers with lower mean and higher standard deviation scores. 

\begin{figure}[!b]
\centering
\includegraphics[trim={0 0 0 0},clip,scale=.15]{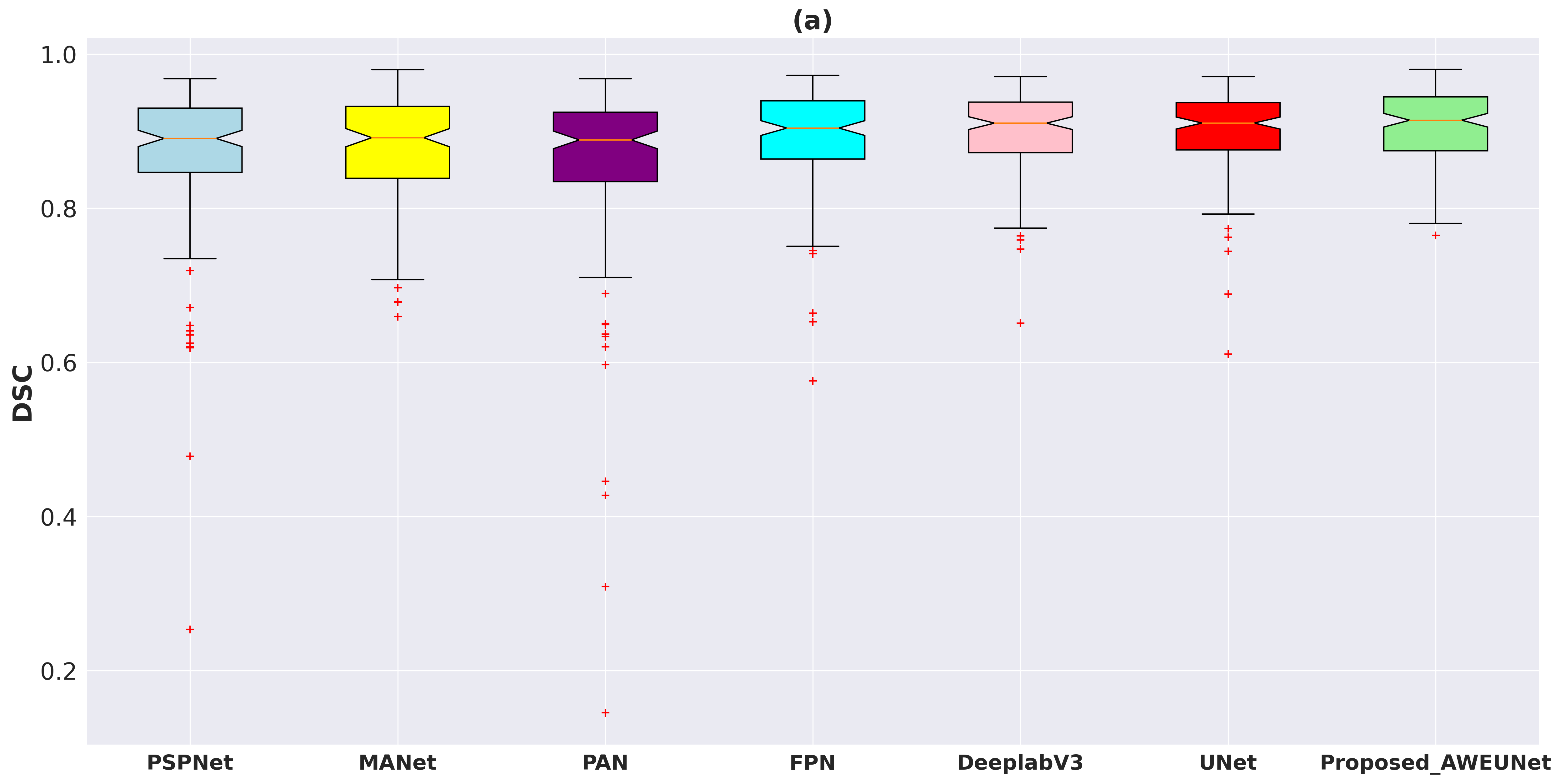}
\includegraphics[trim={0 0 0 0},clip, scale=.15]{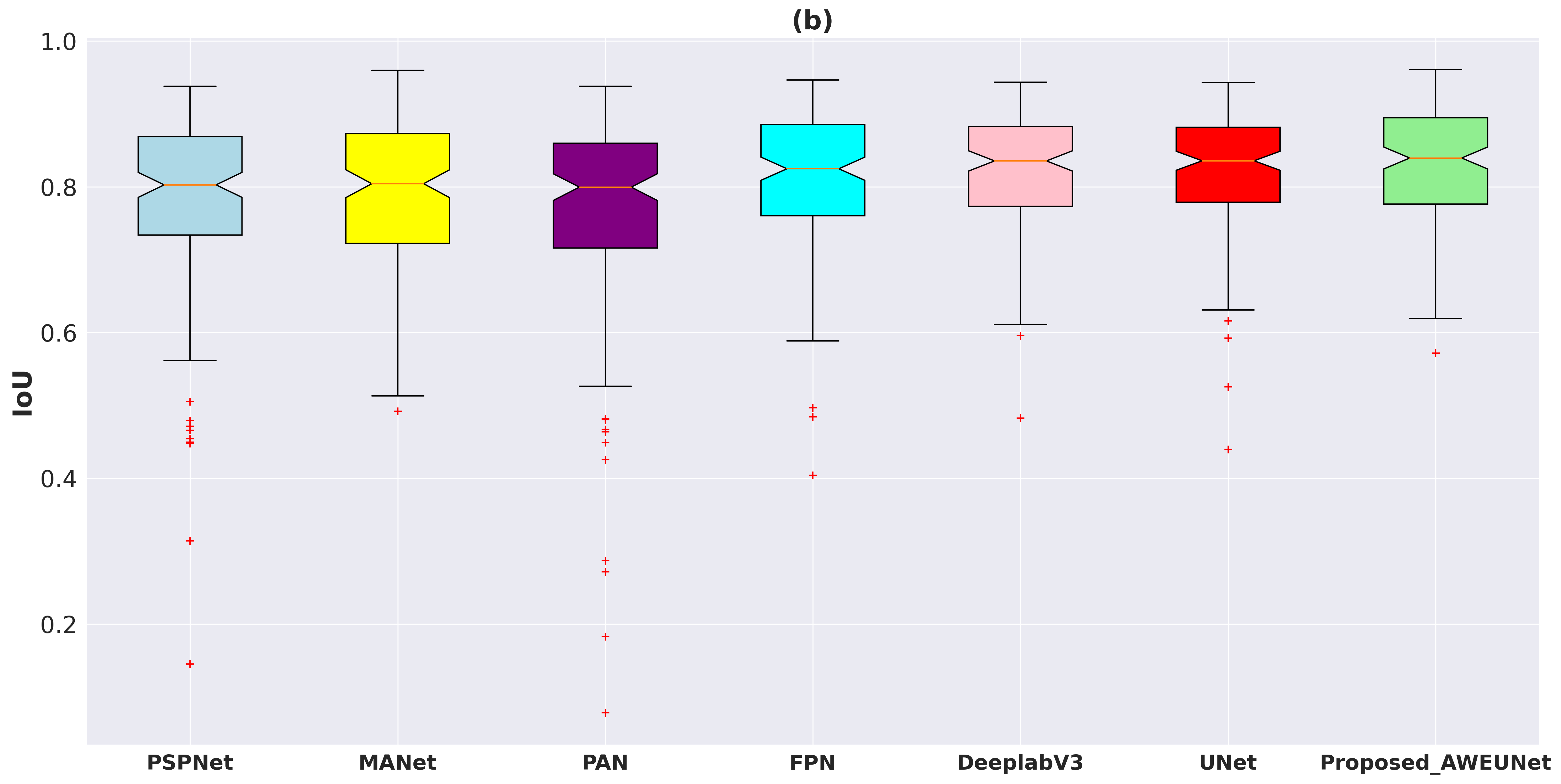}
\caption{Boxplots of (a) Dice coefficient (DSC) and (b) Intersection over Union (IoU) scores for all test samples of LIDC-IDRI lung nodule segmentation dataset. Different boxes indicate the score ranges of several methods; the red line inside each box represents the median value; and all values outside the whiskers are considered as outliers, which are marked with the (+) symbol.}
\label{fig_dsc_iou_lidc}
\end{figure}

\begin{figure}[!t]
\centering
\includegraphics[trim={0 0 0 0},clip,scale=.28]{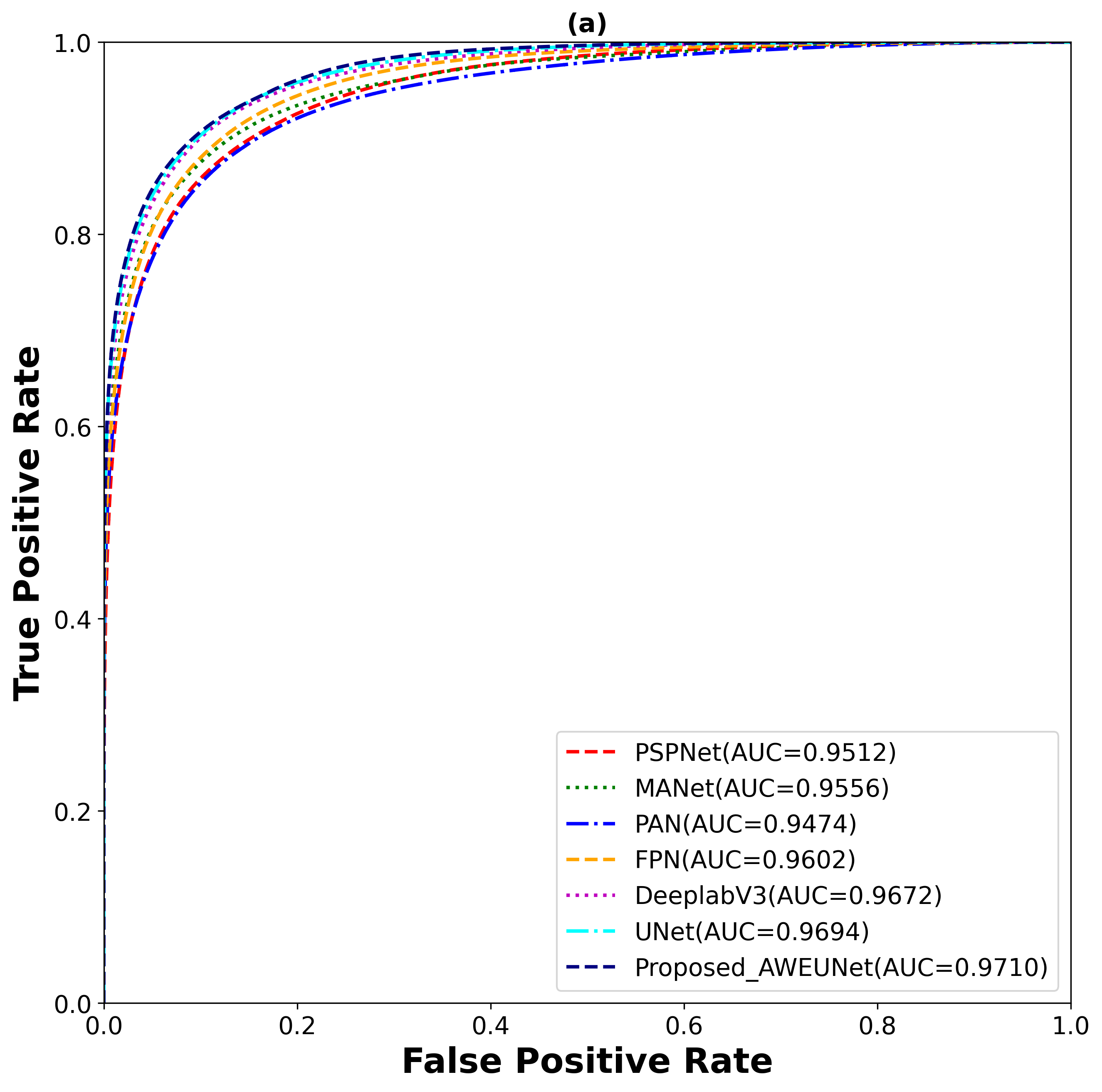}
\includegraphics[trim={0 0 0 0},clip, scale=.28]{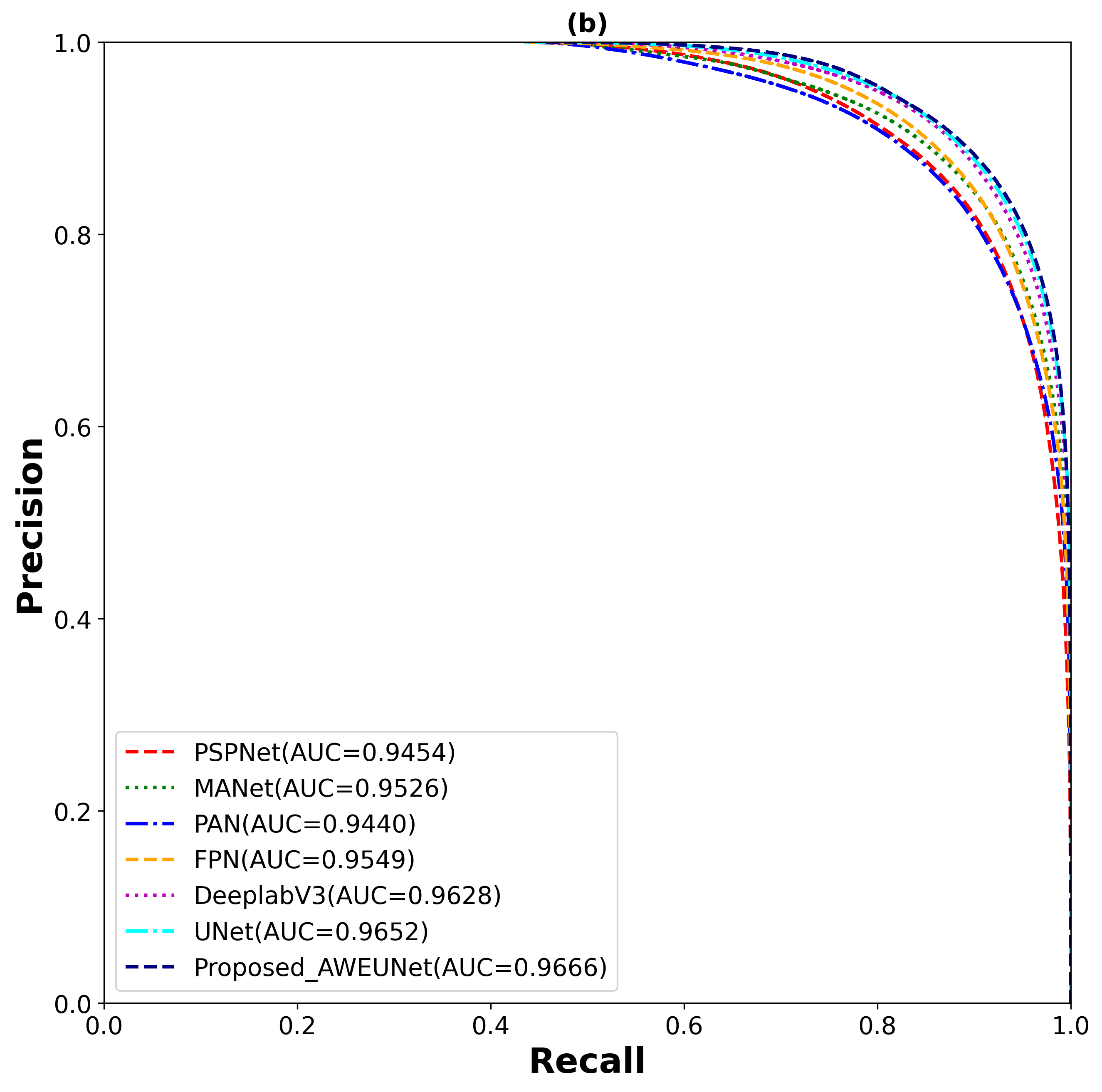}
\caption{The (a) ROC and (b) PR curve for for all test samples of LUNA16 lung nodule segmentation dataset.}
\label{fig_roc_pr_luna16}
\end{figure}

Furthermore, to predict the probability of the binary segmented masks, the ROC and PR curves were constructed as shown in Figure~\ref{fig_roc_pr_luna16}. Using the LUNA16 test set, the proposed AWEU-Net yields the highest AUC and PR of $97.10\%$, and $96.66\%$, respectively among the seven segmentation models tested.  

\begin{figure}[!t]
\centering
\includegraphics[trim={0 0 0 0},clip,scale=.28]{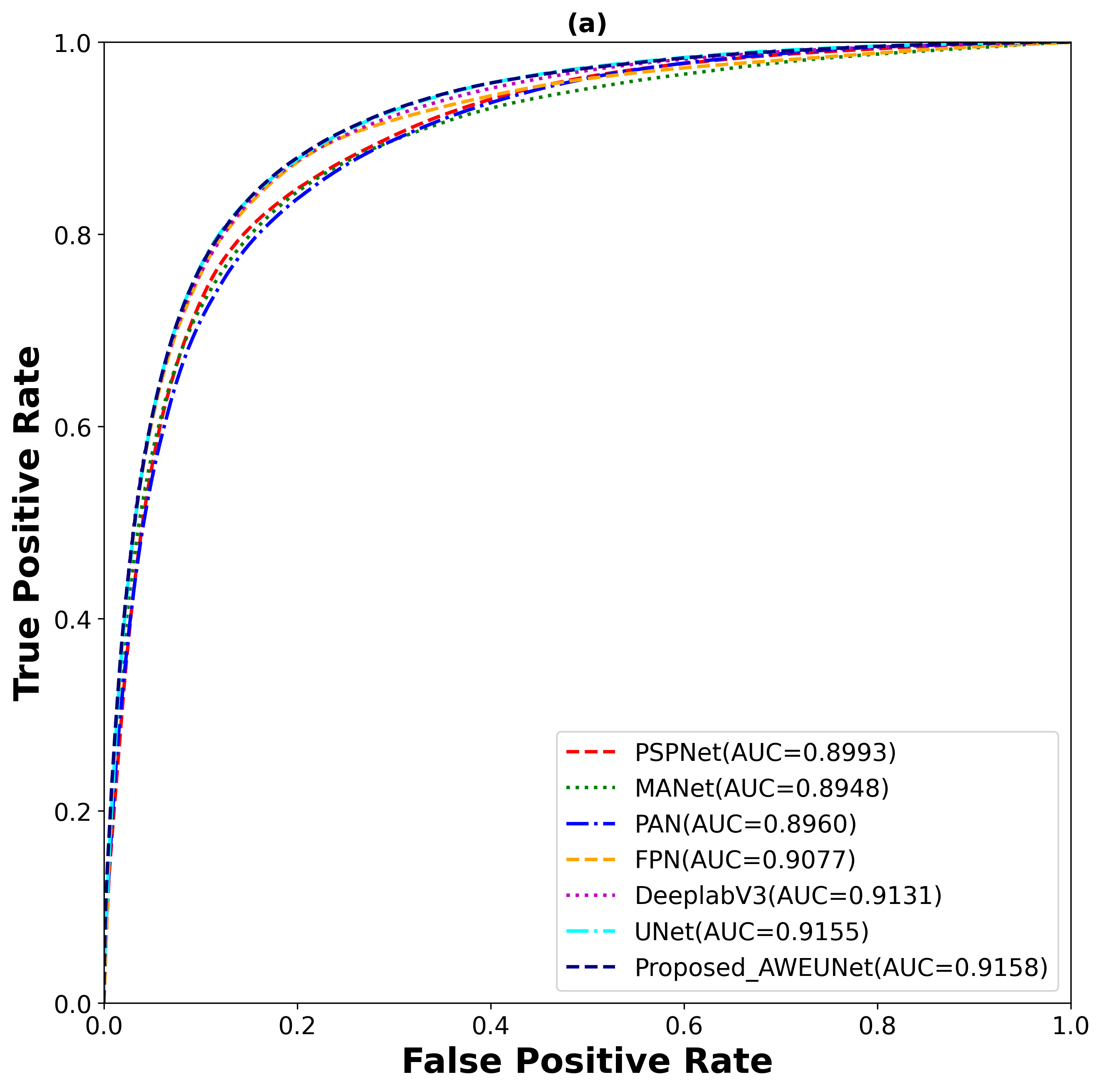}
\includegraphics[trim={0 0 0 0},clip, scale=.28]{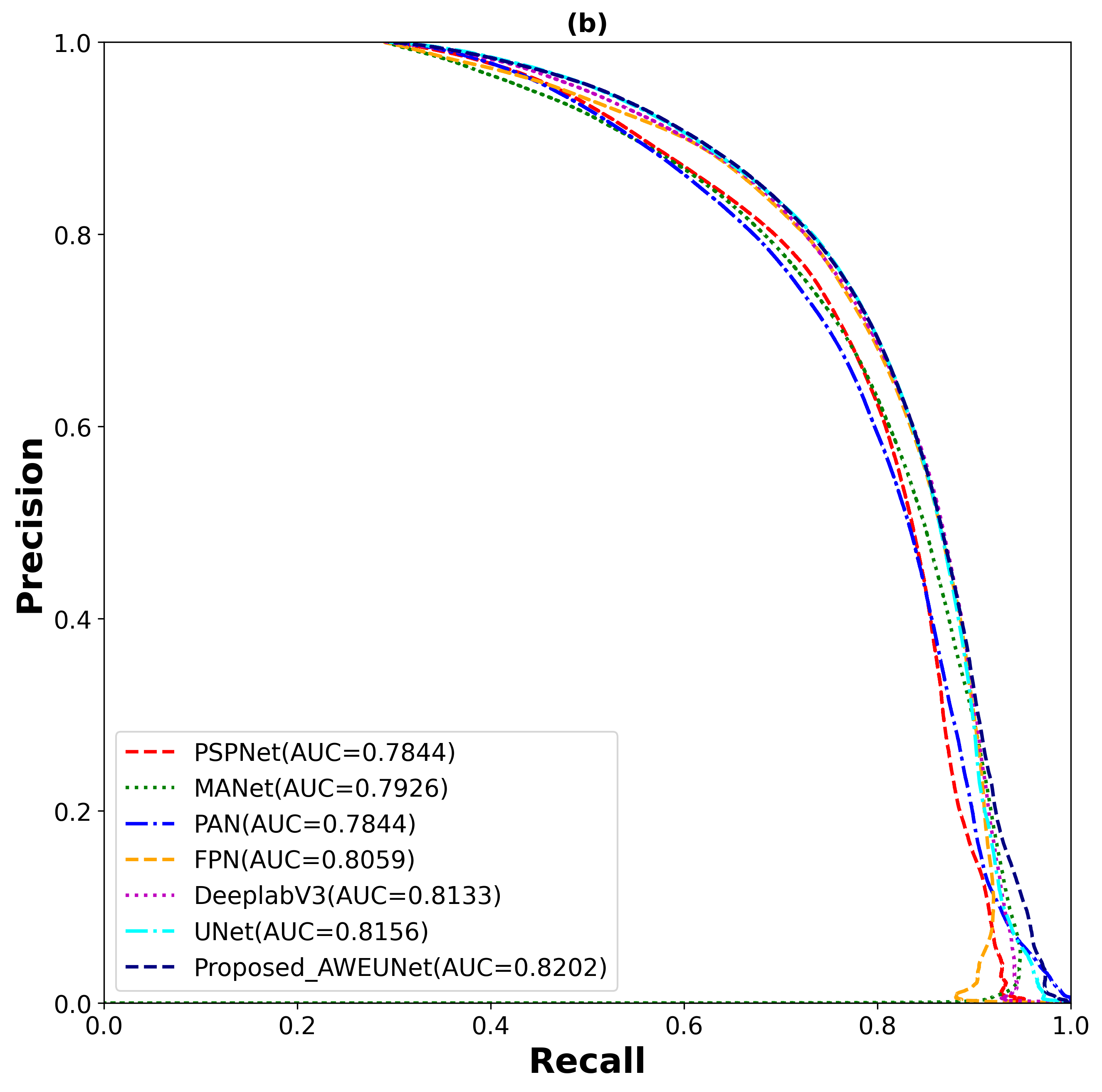}
\caption{The (a) ROC and (b) PR curve for for all test samples of LIDC-IDRI lung nodule segmentation dataset.}
\label{fig_roc_pr_lidc}
\end{figure}

On the other hand, AWEU-Net outperforms all the tested models in terms of all evaluation metrics on the LIDC-IDRI dataset. The proposed model yields the ACC, SEN, SPE, DSC, and IoU scores of 94.66\%, 90.84\%, 96.41\%, 90.35\%, and 83.21\%, respectively. Its improved 0.3\%, 1.16\%, 0.06\%, 0.48\%, and 1.21\% of ACC, SEN, SPE, DSC, and IoU scores from the original U-Net. Again, the box plots of DSC and IoU scores of the LIDC-IDRI  dataset to compare the models performance is displayed in Figure~\ref{fig_dsc_iou_lidc}. Likewise, the proposed AWEU-Net highest DSC and IoU mean scores and the small standard deviation with only one outlier. The proposed model achieved the AUC of ROC and PR on the LIDC-IDRI test dataset are 91.58\%, and 82.02\%, respectively shown in Figure~\ref{fig_roc_pr_lidc}.

\begin{figure}[!t]
\centering
\includegraphics[width=0.7\textwidth]{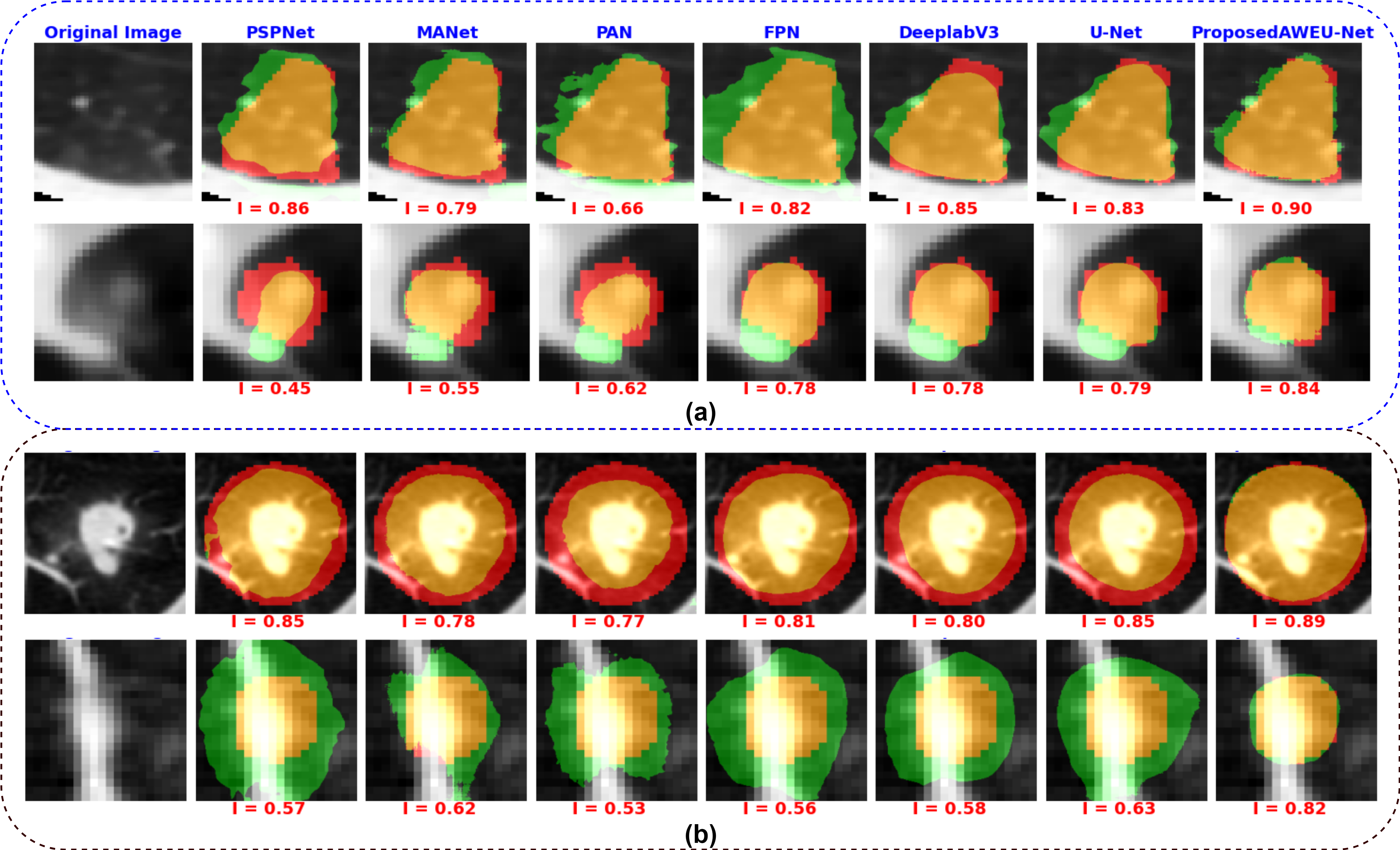}
\caption{Examples of segmentation results by different state-of-the-art models (a) segmentation results on LIDC-IDRI dataset and (b) segmentation results on LUNA16 dataset. The colors of segmentation visualization results are presented as follows: TP (orange), TN (background), FP (green), and FN (red).}
\label{fig_segmentation_examples}
\end{figure}

Finally, a qualitative comparison of the segmentation results of the AWEU-Net and the six segmentation models is shown in Figure~\ref{fig_segmentation_examples}. The segmentation results of the input nodule ROIs of CT images with a variety of difficult levels: illumination variations, irregular shape and boundary of the nodule regions were presented. As shown in Figure~\ref{fig_segmentation_examples}, four examples from the two datasets along with the ground truth and the predicted mask of the six tested models were compared to the proposed AWEU-Net model. AWEU-Net provides segmentation results very close to the ground truth with an average similarity of $> 86\%$ (True Positive (TP)). Our segmentation method also provides the lowest degrees of False Negative (FN) and False Positive (FP) compared to the rest of the models. The AWEU-Net model yields regular borders compared to PSPNet, MANet, FPN, since our model strives for higher accuracy on nodule region boundaries. The resulting segmentation of the six tested models may significantly differ from the ground truth in some cases, e.g., the second example of the LUNA16 dataset.

\section{Conclusions}
\label{conclusions}

This article proposed a reliable system for lung nodule detection and segmentation. The system contains two deep learning models. Firstly, the Optimized Faster R-CNN model~\cite{ren2016faster} trained with lung CT scan images was used for detecting the nodule region in a CT image as an initial step. Secondly, a segmentation model, AWEU-Net, was proposed for segmenting the nodule boundaries of the detected nodule region. The proposed segmentation model, AWEU-Net includes PAWE and CAWE blocks to improve the segmentation performance.  Compared to the state-of-the-art models, the proposed  AWEU-Net yields the best segmentation accuracy with DSC and IoU scores of $89.79\%$, $90.35\%$, and $82.34\%$, $83.21\%$ on the LUNA16 and LIDC-IDRI datasets, respectively. In future work, we will develop a comprehensive end-to-end nodule segmentation system and it will be able to classify and grade the nodule malignancy.

\vspace{6pt} 


\authorcontributions{Conceptualization, S.F.B and M.M.K.S.;
methodology, S.F.B and M.M.K.S.; software, S.F.B and M.M.K.S.; 
validation, S.F.B and M.M.K.S.; formal analysis, S.F.B and M.M.K.S.; investigation, S.F.B and M.M.K.S.; resources, S.F.B and M.M.K.S.; data curation, S.F.B and M.M.K.S.; writing—original draft preparation, S.F.B and M.M.K.S.; writing—review and editing, S.F.B , M.M.K.S , M.A.-N. and H.A.R.; visualization, S.F.B and M.M.K.S.; supervision,  M.A.-N. , D.P. and H.A.R; project administration, M.A.-N. , D.P. and H.A.R.; funding acquisition, M.A.-N. , D.P. and H.A.R.; 
All authors have read and agreed to the published version of the manuscript}

\funding{Not applicable}

\institutionalreview{The study was approved by URV.}

\informedconsent{Not applicable.}

\dataavailability{The samples used are publicly available.} 

\acknowledgments{The Spanish Government partly supported this research through project PID2019-105789RB-I00.}

\conflictsofinterest{All authors declare no conflict of interest.}

\abbreviations{The following abbreviations are used in this manuscript:\\

\noindent 
\begin{tabular}{@{}ll}
CAD & Computer-aided Diagnosis \\
CADe & Computer-aided Detection \\
CASe & Computer-aided Segmentation \\
CT & Computed Tomography \\
AI & Artificial Intelligence \\
GMM & Gaussian Mixture Model \\
CNNs & Convolutional Neural Networks \\
R-CNN & Region-based Convolutional Neural Network \\
ROI &  Region of Interest (ROI) \\
RPN & Region Proposal Network \\
FPN & Feature Pyramid Network \\
ReLU & Rectified Linear Unit  \\
SGD & Stochastic Gradient Descent \\
BCE & Binary Cross-Entropy \\
Dice & Dice Coefficient \\
IoU & Intersection Over Union \\
ROC & Receiver Operating Characteristic \\
AUC & Area Under the Curve \\
ACC & Accuracy \\
SEN & Sensitivity\\
SPE & Specificity \\
GPU & Graphics Processing Unit \\
GB & Gigabytes \\
DL & Deep Learning \\
\end{tabular}}

\end{paracol}
\reftitle{References}

\externalbibliography{yes}
\bibliography{my_bib}

\begin{thebibliography}{999}

\bibitem[can()]{cancer2020}
Cancer.
\newblock \url{https://www.who.int/news-room/fact-sheets/detail/cancer/},
  accessed on 08.08.2021.

\bibitem[fac()]{factsheets2020}
World Lung Cancer Day 2020 Fact Sheet.
\newblock
  \url{https://www.chestnet.org/newsroom/chest-news/2020/07/world-lung-cancer-day-2020-fact-sheet/},
  accessed on 08.08.2021.

\bibitem[Team(2011)]{national2011reduced}
Team, N.L.S.T.R.
\newblock Reduced lung-cancer mortality with low-dose computed tomographic
  screening.
\newblock {\em New England Journal of Medicine} {\bf 2011}, {\em
  365},~395--409.

\bibitem[Callister \em{et~al.}(2015)Callister, Baldwin, Akram, Barnard, Cane,
  Draffan, Franks, Gleeson, Graham, Malhotra, et~al.]{callister2015british}
Callister, M.; Baldwin, D.; Akram, A.; Barnard, S.; Cane, P.; Draffan, J.;
  Franks, K.; Gleeson, F.; Graham, R.; Malhotra, P.; others.
\newblock British Thoracic Society guidelines for the investigation and
  management of pulmonary nodules: accredited by NICE.
\newblock {\em Thorax} {\bf 2015}, {\em 70},~ii1--ii54.

\bibitem[scc()]{sccaprotontherapy2020}
Seattle Cancer Care Alliance Proton Therapy Center.
\newblock
  \url{https://www.sccaprotontherapy.com/cancers-treated/lung-cancer-treatment},
  accessed on 08.08.2021.

\bibitem[Wu \em{et~al.}(2018)Wu, Zhou, Wang, and Wang]{wu2018joint}
Wu, B.; Zhou, Z.; Wang, J.; Wang, Y.
\newblock Joint learning for pulmonary nodule segmentation, attributes and
  malignancy prediction.
\newblock  2018 IEEE 15th International Symposium on Biomedical Imaging (ISBI
  2018). IEEE,  2018, pp. 1109--1113.

\bibitem[Aresta \em{et~al.}(2019)Aresta, Jacobs, Ara{\'u}jo, Cunha, Ramos, van
  Ginneken, and Campilho]{aresta2019iw}
Aresta, G.; Jacobs, C.; Ara{\'u}jo, T.; Cunha, A.; Ramos, I.; van Ginneken, B.;
  Campilho, A.
\newblock iW-Net: an automatic and minimalistic interactive lung nodule
  segmentation deep network.
\newblock {\em Scientific reports} {\bf 2019}, {\em 9},~1--9.

\bibitem[Keetha \em{et~al.}(2020)Keetha, Annavarapu, et~al.]{keetha2020u}
Keetha, N.V.; Annavarapu, C.S.R.; others.
\newblock U-Det: A Modified U-Net architecture with bidirectional feature
  network for lung nodule segmentation.
\newblock {\em arXiv preprint arXiv:2003.09293} {\bf 2020}.

\bibitem[Tang \em{et~al.}(2019)Tang, Zhang, and Xie]{tang2019nodulenet}
Tang, H.; Zhang, C.; Xie, X.
\newblock Nodulenet: Decoupled false positive reduction for pulmonary nodule
  detection and segmentation.
\newblock  International Conference on Medical Image Computing and
  Computer-Assisted Intervention. Springer,  2019, pp. 266--274.

\bibitem[Cao \em{et~al.}(2020)Cao, Liu, Song, Hung, Ma, Xu, Jin, and
  Lu]{cao2020dual}
Cao, H.; Liu, H.; Song, E.; Hung, C.C.; Ma, G.; Xu, X.; Jin, R.; Lu, J.
\newblock Dual-branch residual network for lung nodule segmentation.
\newblock {\em Applied Soft Computing} {\bf 2020}, {\em 86},~105934.

\bibitem[Kumar~Singh \em{et~al.}(2021)Kumar~Singh, Abdel-Nasser, Pandey, and
  Puig]{kumar2021lunginfseg}
Kumar~Singh, V.; Abdel-Nasser, M.; Pandey, N.; Puig, D.
\newblock Lunginfseg: Segmenting covid-19 infected regions in lung ct images
  based on a receptive-field-aware deep learning framework.
\newblock {\em Diagnostics} {\bf 2021}, {\em 11},~158.

\bibitem[Jiang \em{et~al.}(2018)Jiang, Hu, Liu, Halpenny, Hellmann, Deasy,
  Mageras, and Veeraraghavan]{jiang2018multiple}
Jiang, J.; Hu, Y.c.; Liu, C.J.; Halpenny, D.; Hellmann, M.D.; Deasy, J.O.;
  Mageras, G.; Veeraraghavan, H.
\newblock Multiple resolution residually connected feature streams for
  automatic lung tumor segmentation from CT images.
\newblock {\em IEEE transactions on medical imaging} {\bf 2018}, {\em
  38},~134--144.

\bibitem[Ronneberger \em{et~al.}(2015)Ronneberger, Fischer, and
  Brox]{ronneberger2015u}
Ronneberger, O.; Fischer, P.; Brox, T.
\newblock U-net: Convolutional networks for biomedical image segmentation.
\newblock  International Conference on Medical image computing and
  computer-assisted intervention. Springer,  2015, pp. 234--241.

\bibitem[Ren \em{et~al.}(2016)Ren, He, Girshick, and Sun]{ren2016faster}
Ren, S.; He, K.; Girshick, R.; Sun, J.
\newblock Faster R-CNN: towards real-time object detection with region proposal
  networks.
\newblock {\em IEEE transactions on pattern analysis and machine intelligence}
  {\bf 2016}, {\em 39},~1137--1149.

\bibitem[Dehmeshki \em{et~al.}(2008)Dehmeshki, Amin, Valdivieso, and
  Ye]{dehmeshki2008segmentation}
Dehmeshki, J.; Amin, H.; Valdivieso, M.; Ye, X.
\newblock Segmentation of pulmonary nodules in thoracic CT scans: a region
  growing approach.
\newblock {\em IEEE transactions on medical imaging} {\bf 2008}, {\em
  27},~467--480.

\bibitem[Tan \em{et~al.}(2013)Tan, Schwartz, and Zhao]{tan2013segmentation}
Tan, Y.; Schwartz, L.H.; Zhao, B.
\newblock Segmentation of lung lesions on CT scans using watershed, active
  contours, and Markov random field.
\newblock {\em Medical physics} {\bf 2013}, {\em 40},~043502.

\bibitem[Farag \em{et~al.}(2013)Farag, Abd El~Munim, Graham, and
  Farag]{farag2013novel}
Farag, A.A.; Abd El~Munim, H.E.; Graham, J.H.; Farag, A.A.
\newblock A novel approach for lung nodules segmentation in chest CT using
  level sets.
\newblock {\em IEEE Transactions on Image Processing} {\bf 2013}, {\em
  22},~5202--5213.

\bibitem[Dai \em{et~al.}(2015)Dai, Lu, Dong, Zhang, and Chen]{dai2015novel}
Dai, S.; Lu, K.; Dong, J.; Zhang, Y.; Chen, Y.
\newblock A novel approach of lung segmentation on chest CT images using graph
  cuts.
\newblock {\em Neurocomputing} {\bf 2015}, {\em 168},~799--807.

\bibitem[Navya and Pradeep(2018)]{navya2018lung}
Navya, K.; Pradeep, G.
\newblock Lung Nodule Segmentation Using Adaptive Thresholding and Watershed
  Transform.
\newblock  2018 3rd IEEE International Conference on Recent Trends in
  Electronics, Information \& Communication Technology (RTEICT). IEEE,  2018,
  pp. 630--633.

\bibitem[Li \em{et~al.}(2020)Li, Li, Liu, Yin, and Zhou]{li2020segmentation}
Li, X.; Li, B.; Liu, F.; Yin, H.; Zhou, F.
\newblock Segmentation of pulmonary nodules using a GMM fuzzy C-means
  algorithm.
\newblock {\em Ieee Access} {\bf 2020}, {\em 8},~37541--37556.

\bibitem[Savic \em{et~al.}(2021)Savic, Ma, Ramponi, Du, and
  Peng]{savic2021lung}
Savic, M.; Ma, Y.; Ramponi, G.; Du, W.; Peng, Y.
\newblock Lung nodule segmentation with a region-based fast marching method.
\newblock {\em Sensors} {\bf 2021}, {\em 21},~1908.

\bibitem[He \em{et~al.}(2016)He, Zhang, Ren, and Sun]{he2016deep}
He, K.; Zhang, X.; Ren, S.; Sun, J.
\newblock Deep residual learning for image recognition.
\newblock  Proceedings of the IEEE conference on computer vision and pattern
  recognition,  2016, pp. 770--778.

\bibitem[Hancock and Magnan(2016)]{hancock2016lung}
Hancock, M.C.; Magnan, J.F.
\newblock Lung nodule malignancy classification using only
  radiologist-quantified image features as inputs to statistical learning
  algorithms: probing the Lung Image Database Consortium dataset with two
  statistical learning methods.
\newblock {\em Journal of Medical Imaging} {\bf 2016}, {\em 3},~044504.

\bibitem[Lin \em{et~al.}(2014)Lin, Maire, Belongie, Hays, Perona, Ramanan,
  Doll{\'a}r, and Zitnick]{lin2014microsoft}
Lin, T.Y.; Maire, M.; Belongie, S.; Hays, J.; Perona, P.; Ramanan, D.;
  Doll{\'a}r, P.; Zitnick, C.L.
\newblock Microsoft coco: Common objects in context.
\newblock  European conference on computer vision. Springer,  2014, pp.
  740--755.

\bibitem[Fu \em{et~al.}(2019)Fu, Liu, Tian, Li, Bao, Fang, and Lu]{fu2019dual}
Fu, J.; Liu, J.; Tian, H.; Li, Y.; Bao, Y.; Fang, Z.; Lu, H.
\newblock Dual attention network for scene segmentation.
\newblock  Proceedings of the IEEE Conference on Computer Vision and Pattern
  Recognition,  2019, pp. 3146--3154.

\bibitem[Quader \em{et~al.}(2020)Quader, Bhuiyan, Lu, Dai, and
  Li]{quader2020weight}
Quader, N.; Bhuiyan, M.M.I.; Lu, J.; Dai, P.; Li, W.
\newblock Weight Excitation: Built-in Attention Mechanisms in Convolutional
  Neural Networks.
\newblock  European Conference on Computer Vision. Springer,  2020, pp.
  87--103.

\bibitem[Armato~III \em{et~al.}(2011)Armato~III, McLennan, Bidaut, McNitt-Gray,
  Meyer, Reeves, Zhao, Aberle, Henschke, Hoffman, et~al.]{armato2011lung}
Armato~III, S.G.; McLennan, G.; Bidaut, L.; McNitt-Gray, M.F.; Meyer, C.R.;
  Reeves, A.P.; Zhao, B.; Aberle, D.R.; Henschke, C.I.; Hoffman, E.A.; others.
\newblock The lung image database consortium (LIDC) and image database resource
  initiative (IDRI): a completed reference database of lung nodules on CT
  scans.
\newblock {\em Medical physics} {\bf 2011}, {\em 38},~915--931.

\bibitem[Setio \em{et~al.}(2017)Setio, Traverso, De~Bel, Berens, Van
  Den~Bogaard, Cerello, Chen, Dou, Fantacci, Geurts,
  et~al.]{setio2017validation}
Setio, A.A.A.; Traverso, A.; De~Bel, T.; Berens, M.S.; Van Den~Bogaard, C.;
  Cerello, P.; Chen, H.; Dou, Q.; Fantacci, M.E.; Geurts, B.; others.
\newblock Validation, comparison, and combination of algorithms for automatic
  detection of pulmonary nodules in computed tomography images: the LUNA16
  challenge.
\newblock {\em Medical image analysis} {\bf 2017}, {\em 42},~1--13.

\bibitem[Paszke \em{et~al.}(2019)Paszke, Gross, Massa, Lerer, Bradbury, Chanan,
  Killeen, Lin, Gimelshein, Antiga, et~al.]{paszke2019pytorch}
Paszke, A.; Gross, S.; Massa, F.; Lerer, A.; Bradbury, J.; Chanan, G.; Killeen,
  T.; Lin, Z.; Gimelshein, N.; Antiga, L.; others.
\newblock Pytorch: An imperative style, high-performance deep learning library.
\newblock {\em Advances in neural information processing systems} {\bf 2019},
  {\em 32},~8026--8037.

\bibitem[Gulcehre \em{et~al.}(2017)Gulcehre, Sotelo, and
  Bengio]{gulcehre2017robust}
Gulcehre, C.; Sotelo, J.; Bengio, Y.
\newblock {A robust adaptive stochastic gradient method for deep learning}.
\newblock  Neural Networks (IJCNN), 2017 International Joint Conference on.
  IEEE,  2017, pp. 125--132.

\bibitem[Kingma and Ba(2014)]{kingma2014adam}
Kingma, D.P.; Ba, J.
\newblock {Adam: A method for stochastic optimization}.
\newblock {\em arXiv preprint arXiv:1412.6980} {\bf 2014}.

\bibitem[Girshick \em{et~al.}(2014)Girshick, Donahue, Darrell, and
  Malik]{girshick2014rich}
Girshick, R.; Donahue, J.; Darrell, T.; Malik, J.
\newblock Rich feature hierarchies for accurate object detection and semantic
  segmentation.
\newblock  Proceedings of the IEEE conference on computer vision and pattern
  recognition,  2014, pp. 580--587.

\bibitem[Girshick(2015)]{girshick2015fast}
Girshick, R.
\newblock Fast r-cnn.
\newblock  Proceedings of the IEEE international conference on computer vision,
   2015, pp. 1440--1448.

\bibitem[Zhao \em{et~al.}(2017)Zhao, Shi, Qi, Wang, and Jia]{zhao2017pyramid}
Zhao, H.; Shi, J.; Qi, X.; Wang, X.; Jia, J.
\newblock Pyramid scene parsing network.
\newblock  Proceedings of the IEEE conference on computer vision and pattern
  recognition,  2017, pp. 2881--2890.

\bibitem[Fan \em{et~al.}(2020)Fan, Wang, Li, and Wang]{fan2020ma}
Fan, T.; Wang, G.; Li, Y.; Wang, H.
\newblock Ma-net: A multi-scale attention network for liver and tumor
  segmentation.
\newblock {\em IEEE Access} {\bf 2020}, {\em 8},~179656--179665.

\bibitem[Li \em{et~al.}(2018)Li, Xiong, An, and Wang]{li2018pyramid}
Li, H.; Xiong, P.; An, J.; Wang, L.
\newblock Pyramid attention network for semantic segmentation.
\newblock {\em arXiv preprint arXiv:1805.10180} {\bf 2018}.

\bibitem[Lin \em{et~al.}(2017)Lin, Doll{\'a}r, Girshick, He, Hariharan, and
  Belongie]{lin2017feature}
Lin, T.Y.; Doll{\'a}r, P.; Girshick, R.; He, K.; Hariharan, B.; Belongie, S.
\newblock Feature pyramid networks for object detection.
\newblock  Proceedings of the IEEE conference on computer vision and pattern
  recognition,  2017, pp. 2117--2125.

\bibitem[Chen \em{et~al.}(2017)Chen, Papandreou, Schroff, and
  Adam]{chen2017rethinking}
Chen, L.C.; Papandreou, G.; Schroff, F.; Adam, H.
\newblock Rethinking atrous convolution for semantic image segmentation.
\newblock {\em arXiv preprint arXiv:1706.05587} {\bf 2017}.

\end{thebibliography}

\end{document}